\documentclass{article}

\IfFileExists{PRIMEarxiv.sty}{%
  \usepackage{PRIMEarxiv}%
}{%
  \newcommand{\And}{\and}%
  \newcommand{\keywords}[1]{\par\noindent\textbf{Keywords: }#1\par}%
}

\usepackage[utf8]{inputenc}
\usepackage[T1]{fontenc}
\usepackage{hyperref}
\usepackage{url}
\usepackage{booktabs}

\usepackage{multirow}
\usepackage{rotating}
\usepackage{amsfonts}
\usepackage{amsmath}
\usepackage{siunitx}

\usepackage{threeparttable}
\usepackage{nicefrac}
\usepackage{microtype}
\usepackage{fancyhdr}
\usepackage{graphicx}
\graphicspath{{figures/}}
\newcommand{\safeincludegraphics}[2][]{%
  \IfFileExists{#2}{\includegraphics[#1]{#2}}{%
    \fbox{\parbox[c][0.18\textheight][c]{0.9\linewidth}{\centering
      Missing figure: \texttt{\detokenize{#2}}}}%
  }%
}

\usepackage{placeins}
\usepackage{makecell}
\usepackage[table,xcdraw]{xcolor}

\usepackage{soul}
\usepackage[normalem]{ulem}
\useunder{\uline}{\ul}{}
\usepackage{longtable}
\usepackage{float}

\title{Deep Generative Crystal Structure Prediction: A Benchmark Study and a Controlled Test of Prototype Dependence}

\author{
Lai Wei\\
Department of Computer Science and Engineering\\
University of South Carolina\\
Columbia, SC 29201 \\
\textit{Current affiliation: Department of Computer Science and Cybersecurity,}\\
\textit{University of North Georgia, Dahlonega, GA 30597} \\
\And
Ying Feng\\
Department of Computer Science and Engineering\\
University of South Carolina\\
Columbia, SC 29201 \\
\And
Rongzhi Dong\\
Department of Computer Science and Engineering\\
University of South Carolina\\
Columbia, SC 29201 \\
\And
Madeline Miklos\\
Department of Computer Science and Engineering\\
University of South Carolina\\
Columbia, SC 29201 \\
\And
Jianjun Hu$^{*}$ \\
Department of Computer Science and Engineering\\
University of South Carolina\\
Columbia, SC 29201 \\
\texttt{jianjunh@cse.sc.edu} \\
\textit{$^{*}$Corresponding author} \\
}

\hypersetup{
colorlinks=true,
allcolors=blue,
}

\begin{document}
\maketitle

\begin{abstract}
Deep generative models are widely reported to enable \emph{de novo} crystal
structure prediction (CSP), but what they add over template substitution is untested. We benchmark the template method TCSPv3 against twelve deep generative
CSP models on T180, 180 Materials Project targets screened against the MP20
training data (confirmed on T46). TCSPv3 is strongest,
with top-1 and top-20 success of 67.8\% and 90.6\%, against 63.9\%
and 72.2\% for the best generative model, CrystalFlow. Thirty-three
targets (18.3\%) are solved at top-1 by some generative model where TCSPv3 fails,
but only 8 remain at top-20, and 98\% of these algorithm-only successes have a
prototype of the same stoichiometry and space group in the MP20 data or the
template library, and even the one exception, Ba$_6$Al$_{10}$, has a ternary parent with the same framework in MP20. To test dependence
causally, we remove whole stoichiometric prototype families from the training
data and retrain CrystalFlow and EquiCSP. Recovery falls by up to 48\% for the
1:2 binary and 1:1:2 ternary families but not for the 1:1 and 1:1:2:6 families. We explain this by a degeneracy relation between
prototypes: a structure can be reached from retained prototypes with more species by
merging species on equivalent sites, whereas the reverse cannot be inferred.
Removing the perovskite frameworks that the 1:1:2:6 targets share then costs EquiCSP five of its eight consensus recoveries on that family.
The 1:1:2:6 targets, mostly double perovskites, retain hundreds of such
five-species parents. Deep generative models thus behave as implicit template methods: they reuse
prototypes, including ones of a different stoichiometry, without being told to, yet they remain
below the explicit template method at every candidate budget.
\end{abstract}


\section{Introduction}

Crystal structure prediction (CSP), the computational determination of stable
atomic arrangements from chemical composition alone, constitutes one of the
most fundamental and enduring challenges in materials science and condensed
matter physics~\cite{woodley2008crystal, oganov2019structure}. The ability to
accurately predict crystal structures \emph{in silico} has far-reaching
implications for accelerating materials discovery across diverse technological
domains, including energy storage and conversion~\cite{sendek2017machine},
heterogeneous catalysis~\cite{norskov2009towards}, microelectronics and
semiconductors~\cite{choudhary2020recent}, pharmaceutical
polymorphism~\cite{price2018computed}, and emerging quantum
materials~\cite{zunger2018inverse}. Yet despite decades of continuous effort, CSP
remains exceptionally difficult due to the combinatorial explosion of possible
atomic configurations and the highly complex, rugged potential energy landscape
that governs crystalline stability~\cite{wales2003energy}.

CSP is difficult for several reasons. The number of possible atomic arrangements
grows exponentially with system size and with compositional complexity, which
produces a huge configuration space that must be explored
efficiently~\cite{curtarolo2013aflow}. The potential energy surface of
crystalline materials is characterized by numerous local minima separated by
high energy barriers, making it exceedingly difficult to locate the global
minimum corresponding to the thermodynamically stable ground
state~\cite{wales2003energy}. Lastly, the accurate evaluation of energetic
stability requires quantum mechanical calculations, most commonly density
functional theory (DFT), which impose severe computational constraints on the
extent and magnitude of the systems that can be explored~\cite{curtarolo2012high}.

Traditional approaches to CSP have predominantly relied on global optimization
strategies in conjunction with first-principles calculations. Evolutionary
algorithms~\cite{oganov2006crystal, wang2010crystal}, particle swarm
optimization~\cite{wang2010crystal}, basin hopping~\cite{wales1997global}, and
minima hopping~\cite{goedecker2004minima} methods explicitly search the
potential energy surface by iteratively proposing candidate structures and
evaluating their energies. These methods have demonstrated remarkable success,
including the discovery of novel materials under extreme
conditions~\cite{oganov2019structure, ma2017transparent}, but their reliance on
expensive DFT calculations severely limits their applicability to small unit
cells and simple compositions~\cite{jain2013commentary}.

Two data-driven alternatives have since emerged, and the relationship between
them is the subject of this paper. The first is template- or prototype-based
substitution~\cite{griesemer2021high, mehl2017aflow}, which exploits the
observation that many crystal structures belong to a finite set of structural
prototypes; by performing elemental substitution on known templates, these
methods can produce candidates for unexplored compositions. They are
computationally efficient and often remarkably accurate, but are explicitly
bounded by the coverage of existing structural
databases~\cite{wei2026crystal,wei2024cspbench}: they cannot generate topological motifs
absent from their template libraries, and this limitation is transparent by
construction.

The second is deep generative modeling. Rather than carrying out an explicit search of the
energy landscape, generative CSP models learn to sample directly from the
distribution of thermodynamically stable structures encoded in materials
databases~\cite{xie2021crystal, jiao2023crystal, merchant2023scaling}. Early
efforts used latent-variable formulations, notably the Crystal Diffusion
Variational Autoencoder (CDVAE)~\cite{xie2021crystal} and its
composition-conditioned extension cond-CDVAE~\cite{luo2024deep}. The field then shifted
decisively toward diffusion and flow-based frameworks: DiffCSP~\cite{jiao2023crystal}
jointly diffuses lattice parameters and fractional coordinates under periodic
E(3) equivariance; GemsDiff~\cite{klipfel2024vector} introduces hierarchical
generation with explicit space-group prediction; MatterGen~\cite{zeni2023mattergen}
scales score-based diffusion with property-conditioned adapter fine-tuning;
CrystalFlow~\cite{luo2025crystalflow} replaces stochastic denoising with
deterministic flow matching; and SymmCD~\cite{levy2025symmcd} and
EquiCSP~\cite{lin2024equivariant} emphasize explicit crystallographic symmetry
constraints. Autoregressive and sequential methods such as
Uni-3DAR~\cite{lu2025uni}, manifold random-walk formulations such as
CrystalGRW~\cite{tangsongcharoen2025crystalgrw}, text-conditioned models such
as TGDMat~\cite{das2025periodic}, and stochastic-interpolant frameworks such as
OMatG~\cite{hollmer2025open} also broaden the architectural landscape.

The central claim motivating this literature is that generative models perform
\emph{de novo} structure generation, and therefore escape the coverage limits
that bound template substitution. This claim is rarely tested directly.
Standard evaluation reports aggregate match rates against held-out structures,
which cannot distinguish a model that has learned transferable structural
principles from one that has learned to interpolate among its training prototypes.
Recent studies have begun to examine this gap observationally. Most relevant here, Negishi and Walsh~\cite{negishi2026substitution} classify generated crystals as training duplicates, substitution-derived structures, or unmatched by either
criterion, and report that 81–92\% of chemically valid metastable generated
crystals fall into the first two categories, with the effect strongest in
high-symmetry crystal systems. Complementary distributional novelty metrics
have been proposed~\cite{lemat2025genbench, tnovd2025}, and physics-informed
conditioning has been explored as a route to steering generation away from
dominant training motifs~\cite{vasylenko2025physics}.

These analyses establish a strong association between generated structures and
substitution-accessible ones, but they are observational: they characterize
what models produce, not what models require. An intervention is needed to
establish dependence. If a model’s success on a composition is caused by the
presence of related prototypes in its training data, then removing those
prototypes and retraining should degrade performance on that composition; if
the model has acquired transferable structural principles, then it should not.

In this work we perform exactly this intervention, alongside a standardized
benchmark for determining which model is worth intervening in. Our
contributions are as follows.

\textbf{(1) A standardized benchmark that covers all paradigms.} We evaluate
deep generative models of several paradigms and the template baseline TCSPv3
under identical structure-matching, symmetry and consensus criteria on T180, a
set of 180 targets screened to have no overlap with the MP20 training and
validation data, and report top-1, top-5 and top-20 match rates separately.
The ordering is confirmed on T46. Template
substitution attains the highest success rate, 67.8\% at top-1 and 90.6\% at
top-20 against 63.9\% and 72.2\% for the best generative model (both CrystalFlow), and its margin
grows from 7 targets at top-1 to 33 at top-20.

\textbf{(2) A quantified limit on what generation adds beyond the template.}
Decomposing each generative model's predictions against the template baseline
shows that the large majority of its correct predictions are also recovered by
template substitution. Only 33 of 180 targets (18.3\%) are solved at top-1 by
some generative model where TCSPv3 fails, and 8 (4.4\%) at top-20. We further
show that even this margin cannot be interpreted as template-independence: 97\%
of these algorithm-only successes have a prototype of the same stoichiometry and
space group in the MP20 training data or in the template library. Judged by
the \emph{de novo} standard under which these methods are presented, their
marginal contribution over a much cheaper substitution baseline is modest.

\textbf{(3) A controlled prototype-ablation experiment explaining that limit.}
We remove all training structures belonging to a given stoichiometric prototype
family, retrain a generative model from scratch, and evaluate the targets of the
removed families. We do this for CrystalFlow and EquiCSP on four families of T180
targets; in the eight arms, between 0 and 48 percent of the targets
that the full-data model had recovered were lost, with the largest losses for the
1:2 binary and 1:1:2 ternary families.
A follow-up removal of the perovskite frameworks tests the degeneracy explanation directly (Section~\ref{sec:ablation}).
This establishes prototype dependence causally rather than by association and
is, to our knowledge, the first such intervention reported for generative CSP. It
supplies the mechanism behind contribution (2): generative models largely
duplicate the reach of substitution because they draw, to a significant degree, on
the same prototypes.

\textbf{(4) A quantified residual and its likely source.} Some structures remain correctly
predicted when no prototype of their own stoichiometry is available, or, in the ablation, after
their entire prototype family is removed. On T180 the clearest case is Ba$_6$Al$_{10}$, solved at
top-1 by four of the twelve generative models and missed by TCSPv3 at top-20, with no matching
prototype of the same stoichiometry and space group in either corpus. The MP20 data do,
however, contain a ternary structure with the same framework, Ba$_3$Al$_3$Ga$_2$, from which
Ba$_3$Al$_5$ follows by merging two species on equivalent sites. We argue that this
degeneracy relation, in which a prototype with more species serves a query with fewer, explains
much of the residual: generative models use templates of a different stoichiometry without being
told to, which substitution does only with special design. Whether any part of the residual
is independent of such parents is not established, and its size varies with the family.

\section{Methods}
\label{sec:methods}

\subsection{Overview of evaluated algorithms}

Table~\ref{tab:csp_methods} summarizes the methods evaluated in this study. We
group them by generative mechanism rather than by publication lineage: (i)
template retrieval and substitution, (ii) latent-variable models, (iii)
diffusion, flow-matching and stochastic-interpolant models, and (iv)
autoregressive or otherwise sequential constructions. We note explicitly that
several models were not originally proposed as CSP methods and required
adaptation; these are marked, and the adaptation is described in
Section~\ref{sec:adaptation}.

\begin{table*}[ht]
\centering
\caption{Methods evaluated in this study. TMP = template retrieval and
substitution; LVA = latent-variable/autoencoder; DFF = diffusion, flow-matching
or stochastic interpolant; SEQ = autoregressive or sequential construction.
$^\dagger$ denotes a model originally proposed for unconditional crystal
generation and adapted here to the CSP setting by conditioning on target
composition (Section~\ref{sec:adaptation}).}
\label{tab:csp_methods}
\small
\setlength{\tabcolsep}{6.5pt}
\renewcommand{\arraystretch}{1.1}
\begin{tabular}{lllll}
\toprule
\textbf{Category} & \textbf{Method} & \textbf{Year} & \textbf{Core mechanism} & \textbf{Code} \tabularnewline
\midrule
TMP & TCSPv3 \cite{tcspv3} & 2026 & Template retrieval (embedding and periodic-number routes) with soft oxidation-state scoring and radius-aware substitution & \href{https://github.com/usccolumbia/cspbenchmark}{\textcolor{blue}{Link}} \\[3pt]

LVA & CDVAE$^\dagger$ \cite{xie2021crystal} & 2021 & VAE latent space with diffusion decoder & \href{https://github.com/txie-93/cdvae}{\textcolor{blue}{Link}} \\
LVA & cond-CDVAE \cite{luo2024deep} & 2024 & Composition-conditioned CDVAE & \href{https://github.com/ixsluo/cond-cdvae}{\textcolor{blue}{Link}} \\[3pt]

DFF & DiffCSP \cite{jiao2023crystal} & 2023 & Joint lattice/fractional-coordinate diffusion, periodic E(3) & \href{https://github.com/jiaor17/DiffCSP}{\textcolor{blue}{Link}} \\
DFF & EquiCSP \cite{lin2024equivariant} & 2024 & Periodic E(3) diffusion with lattice permutation invariance & \href{https://github.com/EmperorJia/EquiCSP}{\textcolor{blue}{Link}} \\
DFF & GemsDiff \cite{klipfel2024vector} & 2024 & Staged diffusion conditioned on predicted space group & \href{https://github.com/aklipf/gemsdiff}{\textcolor{blue}{Link}} \\
DFF & MatterGen \cite{zeni2023mattergen} & 2023 & Score-based diffusion over lattice, coordinates and types & \href{https://github.com/microsoft/mattergen}{\textcolor{blue}{Link}} \\
DFF & CrystalFlow \cite{luo2025crystalflow} & 2024 & Deterministic flow matching & \href{https://github.com/ixsluo/CrystalFlow}{\textcolor{blue}{Link}} \\
DFF & TGDMat \cite{das2025periodic} & 2025 & Text-conditioned joint diffusion & \href{https://github.com/kdmsit/TGDMat}{\textcolor{blue}{Link}} \\
DFF & SymmCD$^\dagger$ \cite{levy2025symmcd} & 2025 & Diffusion over asymmetric unit and site symmetry & \href{https://github.com/sibasmarak/SymmCD}{\textcolor{blue}{Link}} \\
DFF & OMatG$^\dagger$ \cite{hollmer2025open} & 2025 & Stochastic interpolants unifying diffusion and flow & \href{https://github.com/FERMat-ML/OMatG}{\textcolor{blue}{Link}} \\
DFF & CrystalGRW \cite{tangsongcharoen2025crystalgrw} & 2025 & Geodesic random walk on Riemannian manifolds & \href{https://github.com/trachote/crystalgrw}{\textcolor{blue}{Link}} \\[3pt]

SEQ & Uni-3DAR \cite{lu2025uni} & 2025 & Autoregressive transformer over octree tokens & \href{https://github.com/dptech-corp/Uni-3DAR}{\textcolor{blue}{Link}} \\
\bottomrule
\end{tabular}
\end{table*}

\subsection{TCSPv3 (template baseline)}
\label{sec:tcspv3}
TCSPv3~\cite{tcspv3} is a template-retrieval and substitution method. It keeps the
composition-to-template workflow of TCSP~2.0~\cite{tcsp2}, together with its
BERTOS oxidation-state predictor~\cite{bertos}, Matscholar element
embeddings~\cite{matscholar} and library of 731{,}293 experimentally and
computationally characterized structures pooled from the Materials Project,
Materials Cloud, C2DB and GNoME, but it is not a newly trained model and is not a
de novo generator. Given a query, it proceeds in five steps. (i)~It retrieves
templates whose reduced stoichiometric coefficients match those of the query,
ranking candidate formulas by two routes, the embedding-and-group distance of
TCSP~2.0 and a Villars periodic-number neighborhood adapted from
PNcsp+~\cite{Oran2026PNcspPlus}, and merges the two routes under a fixed
candidate budget. (ii)~It searches coefficient-preserving element assignments
and keeps up to three near-optimal mappings instead of a single one. (iii)~It
replaces the hard oxidation-state filter of TCSP~2.0 by a soft penalty on
mismatch, computed from BERTOS per-species probabilities averaged over up to
four charge-neutral alternatives. (iv)~It penalizes incompatibility between the
radii of the query and template species, using the nearest-shell coordination of
the template. (v)~It substitutes all elements simultaneously, validates the
resulting composition and geometry, and removes duplicate structures with the
pymatgen \texttt{StructureMatcher}, so that distinct polymorphs are kept. Up to
100 valid candidates are pooled, and the 20 with the best native chemical score
(similarity plus radius penalty, ascending) are exported. In the
\emph{full-cell} mode used here, every output has exactly the atom count written
in the query formula.

TCSPv3 also differs from TCSP~2.0 in its leakage control: every template with
the reduced composition of the query is excluded, including by a check against
the source CIF, and the index used here additionally has the MP20 test
structures removed (Section~\ref{sec:testsets}). In the runs reported here
candidates are ranked by the native chemical score and are not relaxed, and the
optional MatterSim energy stage is not used. 
The Supplementary Information gives the details of each stage. Because TCSPv3 is a
template-based method, its dependence on template availability is a design property
rather than a hidden one, which makes it the natural control against which
implicit prototype dependence in generative models can be measured.

\subsection{Deep latent-variable models}

\paragraph{CDVAE.} The Crystal Diffusion Variational
Autoencoder~\cite{xie2021crystal} combines a periodic E(3)-equivariant encoder,
which maps a structure to a latent vector, with a decoder that predicts
composition and lattice from that latent and then refines atomic positions by
annealed Langevin dynamics on a learned score field. CDVAE was proposed for
unconditional generation and for property optimization in latent space, not for
composition-conditioned CSP; our adaptation is described in
Section~\ref{sec:adaptation}.

\paragraph{cond-CDVAE.} cond-CDVAE~\cite{luo2024deep} extends this framework
with explicit conditioning on chemical composition and external pressure,
supplied to both encoder and decoder so that the latent space becomes
chemistry-aware. This makes composition-conditioned generation native rather
than adapted, and we evaluate it as published.

\subsection{Diffusion, flow and interpolant models}

\paragraph{DiffCSP.} DiffCSP~\cite{jiao2023crystal} formulates CSP as joint
denoising diffusion over the lattice matrix and fractional coordinates,
conditioned on composition. Fractional coordinates are diffused with a
wrapped-normal process that respects periodic boundary conditions, while the
lattice is diffused with a standard Gaussian process; the denoiser is a
periodic E(3)-equivariant graph network. Operating in fractional rather than
Cartesian coordinates is central to the method, as it makes periodic
translation invariance exact.

\paragraph{EquiCSP.} EquiCSP~\cite{lin2024equivariant} is likewise a diffusion
model over lattice and fractional coordinates, and its contribution is to
address symmetry invariances that DiffCSP handles only approximately.
Specifically, it enforces lattice permutation invariance and periodic
translation invariance in the training objective, so that structures related by
a change of unit-cell representation or by a global translation are treated as
identical rather than as distinct targets. EquiCSP is one of the two models we subject to ablation in
Section~\ref{sec:ablation}.

\paragraph{GemsDiff.} GemsDiff~\cite{klipfel2024vector} decomposes generation
into stages so that crystallographic validity holds by construction: a
space-group classifier produces a distribution over the 230 groups for a given
composition, a diffusion model generates lattice parameters conditioned on the
sampled group, and an equivariant diffusion network generates atom types and
coordinates for the asymmetric unit only. The full cell is then reconstructed
by applying the group’s symmetry operations.

\paragraph{MatterGen.} MatterGen~\cite{zeni2023mattergen} is a score-based
diffusion model that jointly corrupts and denoises atom types, fractional
coordinates and the lattice, with coordinate noise applied under periodic
boundary conditions and lattice noise biased toward physically reasonable
densities. Its principal architectural contribution is adapter-based
fine-tuning, which allows a large pretrained base model to be steered toward
chemistry, symmetry, or property constraints without retraining. We evaluate
the composition-conditioned setting.

\paragraph{CrystalFlow.} CrystalFlow~\cite{luo2025crystalflow} replaces the
stochastic denoising trajectory with deterministic flow matching: an
equivariant network parameterizes a velocity field transporting a simple prior
to the data distribution, and generation integrates this field with an ODE
solver. Lattice and fractional coordinates are transported jointly. The
deterministic formulation reduces the number of function evaluations required
relative to comparable diffusion samplers.

\paragraph{TGDMat.} TGDMat~\cite{das2025periodic} conditions joint lattice and
coordinate diffusion on a text embedding produced by a pretrained language
model from a natural-language description of the target material, allowing
compositional and property constraints to be expressed in prose rather than as
structured inputs. For CSP evaluation we supply a templated description
encoding the target composition. 

\paragraph{SymmCD.} SymmCD~\cite{levy2025symmcd} generates crystals by
diffusing over the asymmetric unit together with a representation of site
symmetry, rather than over all atoms in the cell, so that symmetry is a modeled
variable rather than an emergent property. The full structure is reconstructed
from the asymmetric unit and the implied symmetry operations. It was proposed
for unconditional generation and is adapted here
(Section~\ref{sec:adaptation}).

\paragraph{OMatG.} OMatG~\cite{hollmer2025open} employs stochastic
interpolants. This framework subsumes both diffusion and flow matching by
specifying an interpolation path between a reference distribution and the data
distribution, with the interpolant coefficients and noise schedule as design
choices. This yields flexibility in trading sample quality against sampling
cost. OMatG targets open-domain generation rather than composition-conditioned
prediction, and its performance here should be read in that light.

\paragraph{CrystalGRW.} CrystalGRW~\cite{tangsongcharoen2025crystalgrw}
formulates generation as a geodesic random walk on the Riemannian manifolds
natural to each crystal degree of freedom: fractional coordinates evolve on the
flat torus, lattice parameters on the manifold of symmetric positive-definite
matrices, and atom types on the probability simplex. A denoiser trained to
reverse this walk recovers structures from noise, optionally conditioned on
properties such as space group. CrystalGRW is \emph{not}
autoregressive and is classified here with the continuous-time generative
models.

\subsection{Autoregressive models}

\paragraph{Uni-3DAR.} Uni-3DAR~\cite{lu2025uni} tokenizes 3D structures via a
coarse-to-fine octree decomposition of space, compresses the resulting token
sequence, and models it with a standard autoregressive transformer, unifying
molecular and crystalline generation in one sequence-modeling framework.
Generation proceeds by next-token prediction over this hierarchical spatial
representation rather than by sequential placement of individual atoms in
Cartesian space.

\subsection{Adaptation of unconditional generators to the CSP setting}
\label{sec:adaptation}

For all three models discussed here—cond-CDVAE, OMatG, and SymmCD—we supplied the target composition as a fixed conditioning input, avoiding unconditional generation with rejection sampling. For cond-CDVAE, we used the published checkpoint (\texttt{condcdvae-4m-v1.0.0}), which was explicitly retrained to accept the target formula as an input during both training and inference. This allows the model to generate fractional coordinates and lattice parameters for the specified atom-type multiset, ensuring the output matches the requested composition by construction. For OMatG, we did not retrain; instead, its \texttt{create\_compositions} utility generates the input directly by repeating the target formula to the desired sample count. The ODE decoder then produces coordinates and lattice for each fixed composition. For SymmCD, we also did not retrain. The released sampling script constructs the atom-type input from the requested formula and generates only the coordinates, lattice, and site symmetry, fixing composition at generation. In all three cases, we did not need acceptance or rejection sampling. Since composition is fixed as an input, the acceptance rate is effectively 100 percent, and no samples are rejected due to composition mismatch. Therefore, the reported success rates for these models directly reflect structural-validity failures, not composition-mismatch failures. Full parameter configurations and conditioning details are further documented in Supplementary Table 2.


\subsection{Test sets}
\label{sec:testsets}

We evaluate on two test sets of experimentally and computationally characterized
structures drawn from the Materials Project~\cite{jain2013commentary}: T180, a
180-structure set that follows the design of the CSPBench test
set~\cite{wei2024cspbench}, and T46, a smaller 46-structure set with the same
category scheme. Both are listed in the supplementary material (T46 in Supplementary Table~9 and T180 in Supplementary Table~10). Every target of both sets has been screened against the MP20
training and validation data, as described below.

\paragraph{Composition of T180.} T180 contains 60 binary, 60 ternary and 60
quaternary targets. Within each arity the targets are stratified into
evaluation categories: easy, medium and hard structures (12 per level and
arity), template-based structures (12 per arity), and polymorphic or
non-polymorphic structures (the remaining 12 per arity: 10, 9 and 11 polymorphs
for binary, ternary and quaternary compositions, respectively, with 2, 3 and 1
non-polymorphic structures filling the rest). The primitive cells contain
between 2 and 28 atoms (median 10) and span 61 space groups and all seven
crystal systems (orthorhombic 76, hexagonal 27, tetragonal 24, trigonal 18,
cubic 17, monoclinic 12, triclinic 6). The easy/medium/hard and template-based
labels are selection proxies carried over from the original benchmark
construction, not newly measured prediction difficulty.

\paragraph{Composition of T46.} T46 contains 13 binary, 15 ternary and 18
quaternary targets in the same category scheme, with primitive cells of 3 to 24
atoms and 31 space groups. It is a stratified small tier and not an independent
cohort: 33 of its 46 targets also belong to T180, and the remaining 13 do not.
Results on T46 should therefore not be pooled with, or counted as independent
confirmation of, results on T180.

\paragraph{Screening against MP20.} The models we retrained, and several of the
published checkpoints we use, are trained on the MP20 dataset. We therefore
audited both test sets against the full MP20 training and validation data
(36{,}183 structures in total). Every target was re-read from its CIF file and
compared with every compositionally compatible MP20 structure in both directions
with the pymatgen \texttt{StructureMatcher} at two tolerance settings, our
evaluation setting (\texttt{ltol}\,=\,0.2, \texttt{stol}\,=\,0.3,
\texttt{angle\_tol}\,=\,5) and the looser setting used in the CDVAE evaluation
(0.3, 0.5, 10), each with and without supercell matching
(\texttt{primitive\_cell}\,=\,True, \texttt{scale}\,=\,True,
\texttt{allow\_subset}\,=\,False). Targets with a match were replaced. In T180, 35 of the 180 targets were replaced by
targets of the same arity and evaluation category, drawn from the experimentally
sourced (\texttt{theoretical}\,=\,False) MPTS52 test split and restricted to at
most 20 sites; the other 145 are unchanged. In T46, 6 of 46 targets were
replaced and 40 are unchanged (the replaced targets are listed in Supplementary Table~7). After replacement, neither set contains an MP20
identifier or a structural match to an MP20 training or validation structure.
The sets also contain no duplicate compositions.

We are explicit about what this audit does and does not establish. It
guarantees zero overlap with the MP20 training and validation data, which is the
training corpus of the models we retrained. It does not establish absence of the
targets from other corpora: the replacement targets come from MPTS52, and the
CrystalGRW checkpoint was trained on a larger corpus whose
per-structure composition is not available to us (the Alex-MP-20 corpus). The Uni-3DAR
CSP checkpoint is trained on MP-20 only, following the authors' released training script, and the MatterGen and GemsDiff
checkpoints were trained by us on the MP-20 training split. Supplementary Table~1 therefore lists a basis
for each method's exposure and reports zero MP20 overlap only for methods whose
training data we control. The replacement targets were screened against the MP20
training and validation structures with the same identifier and structure-match
criteria as the other targets (for the last replacement, mp-3576, at two matcher
settings and with and without supercell matching), and every model we retrained uses the standard MP-20 split,
so the screen covers all of them.

\paragraph{Query formulas and template baseline.} Each method is given the
formula of the primitive cell of the target as the query, with integer atom
counts (for example Co$_2$Te$_2$ for CoTe). For 114 of the 180 T180 targets this
primitive-cell formula differs from the reduced formula, so the task specifies
the cell size as well as the composition and is not composition-only blind
prediction. The primitive-cell formulas were verified against the CIF files
and corrected where needed. TCSPv3 is a template-based method, not
a trained model, so its exposure is assessed through its template library. The
library index used here is the 731{,}293-structure TCSP index with the MP20 test
structures removed, and for every query TCSPv3 skips each template whose reduced
composition equals that of the query, including by a check against the source
CIF, so that no target can be predicted from its own entry. For T180 this excludes
the 180 target compositions (one exclusion list per test set).

\subsection{Evaluation protocol}
\label{sec:protocol}

\paragraph{Structure matching.} A prediction is counted as a match when
Pymatgen's~\cite{ong2013python} \texttt{StructureMatcher}, with
\texttt{ltol}\,=\,0.2, \texttt{stol}\,=\,0.3, \texttt{angle\_tol}\,=\,5,
\texttt{primitive\_cell}\,=\,True, \texttt{scale}\,=\,True and
\texttt{attempt\_supercell}\,=\,False, finds the predicted and reference
structures equivalent. As a sensitivity check, we also report a strict
primitive-cell variant in which both structures are first reduced to their
primitive cells at a tolerance of 0.01~\AA{} and the matcher's own primitive
reduction is disabled. Some original publications adopt looser criteria. CDVAE,
for instance, used \texttt{ltol}\,=\,0.3, \texttt{stol}\,=\,0.5,
\texttt{angle\_tol}\,=\,10~\cite{xie2021crystal}, so our reported rates are
systematically stricter than published values and should not be compared to
them directly.

\paragraph{Space-group agreement and consensus.} Space groups are determined
with a symmetry tolerance of 0.1 for the predicted structure and the
reference, and the reference space group is recomputed with the same tolerance
rather than taken from the test-set metadata, whose space groups are assigned at
0.01. We also report the space-group rate at \texttt{symprec}\,=\,0.01 as a
sensitivity check. The consensus rate counts a target when the same top-1
candidate satisfies both the structure match and the space-group criterion.

\paragraph{Top-$k$ reporting.} Each method returns up to 20 candidates per
target. We report top-1, top-5 and top-20 match rates. Top-1 is the primary
metric, because it corresponds to the decision a practitioner actually faces
when a single candidate is carried forward to DFT, and it is the only setting
in which template retrieval and generative sampling are compared on equal
terms. Top-20 enables comparison with the published literature, where match
rates over 20 samples are the most common report. A target is a success at
rank $k$ if any of the first $k$ ranked candidates matches. Targets for which a
method returns no valid candidate count as failures at every $k$; the number of
such targets, and the number of targets with a partial pool of fewer than 20
valid candidates, is reported for each method in the supplementary material (Supplementary Table~3), so
that incomplete coverage is never mistaken for a negative prediction.

\paragraph{Input specification and candidate validation.} Every method is
given the primitive-cell formula of the target (Section~\ref{sec:testsets}) and
no lattice, coordinates, space group or Wyckoff information. Each exported
candidate is checked against the requested element counts, and candidates that
fail this check are discarded and not scored. Targets larger than a model's training range (four T180 targets have primitive cells of 22 to 28 atoms, above the 20-atom limit of MP20) are retained and labeled as cell-size extrapolation, not dropped. No method receives an override; all are run with their default settings.

\paragraph{Polymorphs.} For compositions with multiple known polymorphs, the
predicted structure is compared against each ground-truth polymorph and the
smallest distance is retained.

\paragraph{Relaxation, ranking and energy evaluation.} Candidates are
post-processed by the pipeline of each method before ranking: generated
structures are relaxed and ranked with the ORB-v3 universal interatomic
potential, CSPML candidates are relaxed with MatterSim, and TCSPv3 ranks by its native chemical score without relaxation. The relaxer therefore differs between method
families, and the top-1 candidate is the one ranked first by that method's own
procedure. For all twelve generative models the export protocol is the same: the
candidates are relaxed with ORB-v3 (500 steps, force threshold 0.01~eV/\AA) and ranked by
ORB-v3 energy per atom, and the 20 best are exported. The MatterSim potential is used
only for the energy distances of the case studies, not for ranking. Energy distances in the case
studies are computed between structures relaxed with the MatterSim
potential~\cite{mattersim}, identically for the predicted and ground-truth
structures, and are labeled ``energy distance'' throughout.

\paragraph{Running parameters and sampling variance.} All algorithms use the
default checkpoints and sampling settings from their publications and official
repositories; no additional hyperparameter tuning was performed. The full
parameter settings and configurations of all algorithms are listed in
Supplementary Table~2. Each method is run once per target with a fixed sampling
configuration, so the rates reported here carry no repetition-based error bars,
and we do not report sampling variance. With 180 targets a single target is
0.56 percentage points, and with 46 targets it is 2.2 points, so differences of
a few points between methods on T46 correspond to one or two structures.

\section{Results}
\label{sec:results}

\subsection{Benchmark performance on T180}
\label{sec:bench180}

\begin{figure}[tbh!]
\centering
\includegraphics[width=\linewidth]{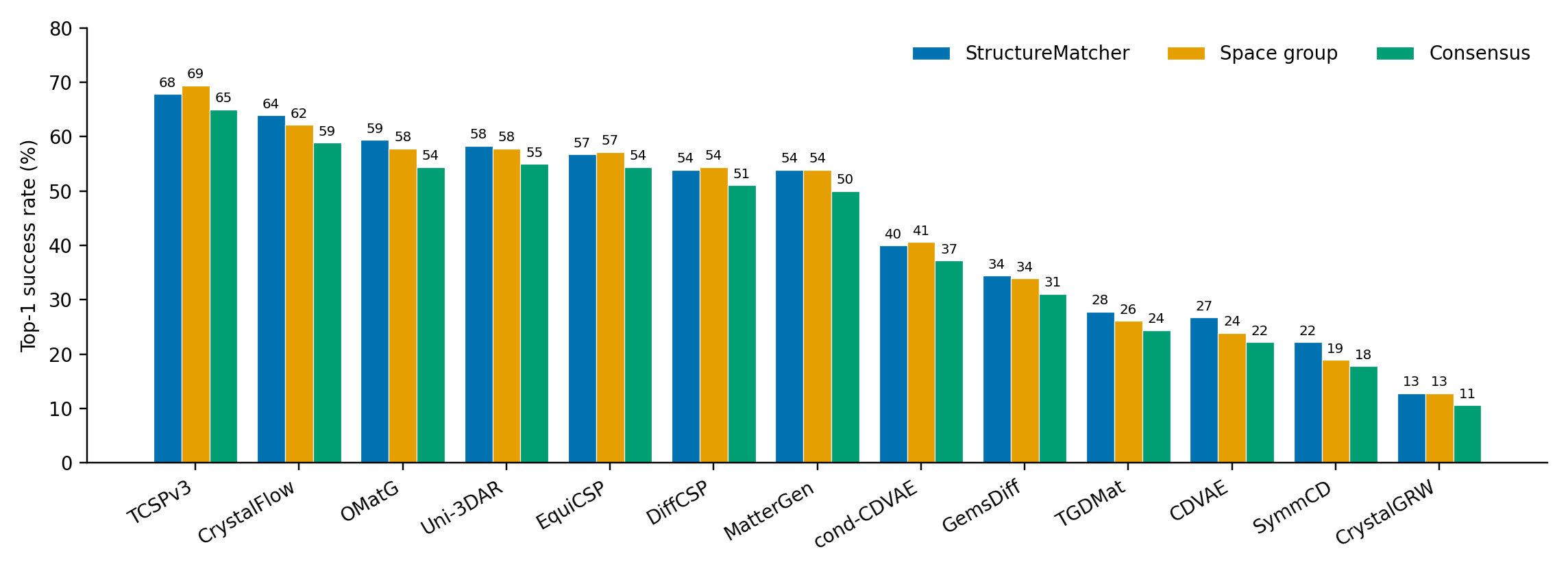}
\caption{Top-1 performance of twelve deep generative CSP algorithms and the
TCSPv3 template baseline on T180. Bars show the StructureMatcher success rate,
the space-group match rate, and the consensus rate, computed with
\texttt{ltol}=0.2, \texttt{stol}=0.3, \texttt{angle\_tol}=5 and a space-group
tolerance of 0.1. Each method was run once per target, so no error bars are
shown.}
\label{fig:succ-180}
\end{figure}

Table~\ref{tab:main180} and Figure~\ref{fig:succ-180} report the top-1, top-5
and top-20 StructureMatcher success rates and the top-1 space-group and
consensus rates of the template baseline TCSPv3 and twelve generative models on
the 180 targets of T180. TCSPv3 achieves the highest top-1
StructureMatcher success rate at 67.8\% (122 of 180 targets), with a
space-group match rate of 69.4\% and a consensus rate of 65.0\%. Among
generative models, CrystalFlow is strongest at 63.9\% (58.9\% consensus), followed
by OMatG at 59.4\% (54.4\%), Uni-3DAR at 58.3\% (55.0\%),
EquiCSP at 56.7\% (54.4\%), and DiffCSP and MatterGen at
53.9\% each (51.1\% and 50.0\% consensus). The older and more specialised models
follow at a distance: cond-CDVAE reaches 40.0\% (37.2\%), GemsDiff 34.4\% (31.1\%),
TGDMat 27.8\% (24.4\%), CDVAE 26.7\% (22.2\%), SymmCD 22.2\% (17.8\%) and CrystalGRW 12.8\%
(10.6\%). The
lead of TCSPv3 over the best generative model is 7 targets (3.9 percentage
points) at top-1, and it grows with the number of candidates: at top-20, TCSPv3
succeeds on 163 targets (90.6\%), against 130 (72.2\%)
for CrystalFlow, the best generative model, and 114 to 124 for the other leading models. The top-1 and
top-20 numbers differ because TCSPv3's candidate pool is more diverse than a
generative model's, and a correct template is often present but not ranked
first: TCSPv3 misses 58 targets at top-1 and recovers 41 of them within its top
20. When TCSPv3's candidates are also relaxed and ranked with ORB-v3, as those of the
generative models are, it solves 130 targets (72.2\%) at top-1 and 159 (88.3\%) at top-5, leads of 15 and 38 targets
over the best generative model, and it keeps the highest space-group (71.1\%) and consensus (70.0\%) rates.

\begin{table}[tb]
\centering
\caption{\textbf{Performance on the 180 targets of T180.} StructureMatcher
success at top-1, top-5 and top-20 (number of targets; percentage of 180 in
parentheses for top-1 and top-20), and top-1 space-group and consensus rates.
Space groups are determined at \texttt{symprec}\,=\,0.1. Consensus requires the
top-1 candidate to satisfy both the StructureMatcher and the space-group
criterion. Methods are sorted by top-1 success. The italic row gives TCSPv3 with its candidates relaxed and ranked with ORB-v3, the same post-processing as the generative models; the other TCSPv3 row uses its native chemical ranking. Targets for which a method
returned no valid candidate count as failures (Uni-3DAR: 1 target; CrystalGRW: 2
targets).}
\label{tab:main180}
\small
\begin{tabular}{lrrrrr}
\toprule
\textbf{Method} & \textbf{Top-1} & \textbf{Top-5} & \textbf{Top-20} & \textbf{Space group} & \textbf{Consensus} \\
\midrule
TCSPv3 & 122 (67.8\%) & 154 & 163 (90.6\%) & 125 (69.4\%) & 117 (65.0\%) \\
\emph{TCSPv3 + ORB-v3 relax} & \emph{130 (72.2\%)} & \emph{159} & \emph{163 (90.6\%)} & \emph{128 (71.1\%)} & \emph{126 (70.0\%)} \\
CrystalFlow & 115 (63.9\%) & 121 & 130 (72.2\%) & 112 (62.2\%) & 106 (58.9\%) \\
OMatG & 107 (59.4\%) & 116 & 124 (68.9\%) & 104 (57.8\%) & 98 (54.4\%) \\
Uni-3DAR & 105 (58.3\%) & 110 & 114 (63.3\%) & 104 (57.8\%) & 99 (55.0\%) \\
EquiCSP & 102 (56.7\%) & 111 & 119 (66.1\%) & 103 (57.2\%) & 98 (54.4\%) \\
DiffCSP & 97 (53.9\%) & 108 & 115 (63.9\%) & 98 (54.4\%) & 92 (51.1\%) \\
MatterGen & 97 (53.9\%) & 110 & 119 (66.1\%) & 97 (53.9\%) & 90 (50.0\%) \\
cond-CDVAE & 72 (40.0\%) & 82 & 86 (47.8\%) & 73 (40.6\%) & 67 (37.2\%) \\
GemsDiff & 62 (34.4\%) & 64 & 66 (36.7\%) & 61 (33.9\%) & 56 (31.1\%) \\
TGDMat & 50 (27.8\%) & 52 & 53 (29.4\%) & 47 (26.1\%) & 44 (24.4\%) \\
CDVAE & 48 (26.7\%) & 52 & 53 (29.4\%) & 43 (23.9\%) & 40 (22.2\%) \\
SymmCD & 40 (22.2\%) & 45 & 49 (27.2\%) & 34 (18.9\%) & 32 (17.8\%) \\
CrystalGRW & 23 (12.8\%) & 24 & 24 (13.3\%) & 23 (12.8\%) & 19 (10.6\%) \\
\bottomrule
\end{tabular}
\end{table}

Three observations stand out. First, CrystalFlow ranks first among the generative models
and EquiCSP fourth. For the ablation (Section~\ref{sec:ablation}) we keep both:
CrystalFlow is the strongest generative model on T180, and EquiCSP is a
second diffusion-based model of lower rank, so the two together show whether the effect of
removing a prototype family depends on the model's rank. Second, the lead of TCSPv3 is concentrated on ternary targets: it
solves 44 of the 60 ternary targets at top-1 (Supplementary Table~5), against at most 38 for any
generative model, whereas on binary targets it is below the best generative model (37
against 41 of 60) and on quaternary targets only slightly ahead (41 against 39). Third, the
gap between structure matching and symmetry agreement that we previously
observed for CrystalFlow is not present here. CrystalFlow matches 63.9\% of
targets and recovers the space group on 62.2\%. Space-group agreement does,
however, depend strongly on the symmetry tolerance for generative models.
Evaluated at \texttt{symprec}\,=\,0.01 (Supplementary Table~4), the top-1 space-group rate of TCSPv3 is
unchanged at 67.8\% (122 targets), whereas it falls to 35.0\% for CrystalFlow
(63), 36.7\% for OMatG (66), 41.1\% for MatterGen (74),
43.9\% for EquiCSP (79), 45.6\% for DiffCSP (82) and 46.7\% for Uni-3DAR (84). Generated structures
are close to the correct symmetric structure but carry coordinate noise that a
tight tolerance resolves as lower symmetry, while template substitution inherits
the exact symmetry of its template. Reporting structure matching and symmetry
agreement together, and stating the tolerance, is therefore necessary.

\subsection{Performance on T46}
\label{sec:bench46}

\begin{figure}[!htb]
\centering
\includegraphics[width=0.95\linewidth]{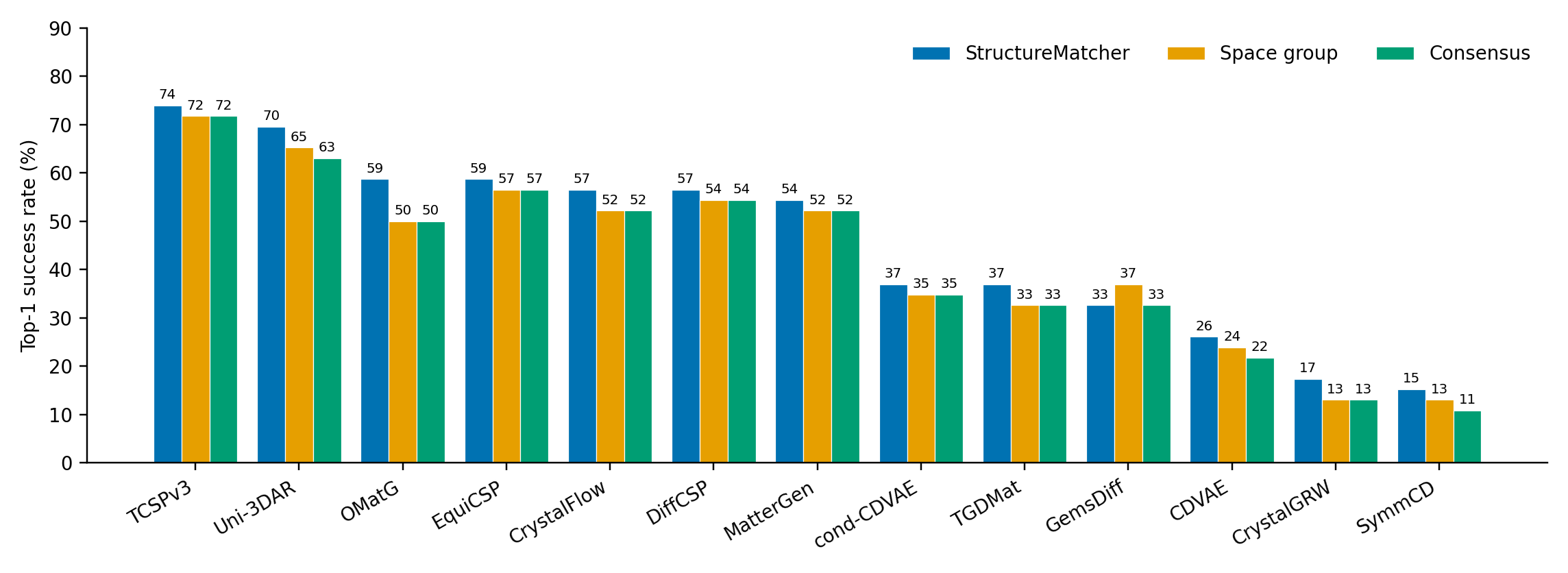}
\caption{Top-1 performance of the 13 methods on T46: StructureMatcher success rate,
space-group match rate and consensus rate, with the same matcher settings as
Figure~\ref{fig:succ-180}. One run per method and target, so no error bars; one of the 46
targets corresponds to 2.2 percentage points.}
\label{fig:succ-46}
\end{figure}

Thirty-three of the 46 T46 targets also belong to T180, and their predictions come
from the T180 campaign; the other 13 targets, which include the 6 targets
replaced in the audit, were run for every method with the same protocol. Table~\ref{tab:t46} and Figure~\ref{fig:succ-46} report the 13 methods of
Section~\ref{sec:bench180}. Because T46 overlaps T180 in 33 targets it is not an independent
confirmation of the T180 ranking, and with 46 targets one target is 2.2 percentage points.

TCSPv3 again leads, with 34 of 46 targets at top-1 (73.9\%) and 41 at top-20 (89.1\%), followed by
Uni-3DAR (32, 69.6\%). The next five generators, from OMatG to
MatterGen, solve 25 to 27 targets (54--59\%) at top-1, a spread of two targets, and their intervals overlap
completely with each other and with TCSPv3, so the order among them
should not be interpreted. The lower half broadly follows the T180 ordering, with TGDMat and GemsDiff and SymmCD and CrystalGRW exchanging places:
cond-CDVAE and TGDMat (17), GemsDiff (15), CDVAE (12), CrystalGRW (8), SymmCD (7). The T46 rate of a method is not
systematically higher or lower than its T180 rate, but individual methods move
by up to ten points (Uni-3DAR 69.6\% against 58.3\% on T180; CrystalFlow 56.5\% against 63.9\%),
which is of the order of the sampling noise (a standard error of about 7 points at $n=46$).

The raw top-1 column shows the same dependence on relaxation as on T180: the raw candidates of cond-CDVAE and CDVAE match only 2 and 1 of the 46 targets, against 17 and 12 after relaxation and ranking, whereas for TCSPv3, which is not relaxed, and for OMatG and EquiCSP the raw and exported counts coincide or differ by at most one target.

\begin{table}[!htb]
\centering
\caption{Match rates on T46 (46 targets; 33 shared with T180, 13 new runs). Top-$k$ counts are targets matched
by one of the first $k$ candidates (percent in parentheses); \emph{SG} is top-1 space-group agreement (symprec 0.1),
\emph{Cons.} the top-1 consensus, \emph{Raw} the top-1 count for the raw, unrelaxed rank-1 candidate (the F0 bucket of the diagnosis; 33 targets from the T180 diagnosis, 13 new), and the last column the 95\% bootstrap interval of the top-1 rate (\%). Targets without a valid
candidate count as misses (\emph{Cov.} gives the number of targets with candidates).}
\label{tab:t46}
\renewcommand{\arraystretch}{1.1}
\small
\begin{tabular}{@{} l r r r r r r r r @{}}
\toprule
\textbf{Method} & \textbf{Cov.} & \textbf{Top-1} & \textbf{Top-5} & \textbf{Top-20} & \textbf{SG} & \textbf{Cons.} & \textbf{Raw} & \textbf{CI} \\
\midrule
TCSPv3 & 46 & 34 (73.9) & 39 (84.8) & 41 (89.1) & 33 & 33 & 34 & 61--85 \\
Uni-3DAR & 45 & 32 (69.6) & 32 (69.6) & 33 (71.7) & 30 & 29 & 30 & 56--83 \\
OMatG & 46 & 27 (58.7) & 28 (60.9) & 30 (65.2) & 23 & 23 & 27 & 46--72 \\
EquiCSP & 46 & 27 (58.7) & 29 (63.0) & 29 (63.0) & 26 & 26 & 27 & 44--72 \\
CrystalFlow & 46 & 26 (56.5) & 27 (58.7) & 31 (67.4) & 24 & 24 & 27 & 41--70 \\
DiffCSP & 46 & 26 (56.5) & 28 (60.9) & 31 (67.4) & 25 & 25 & 27 & 41--70 \\
MatterGen & 46 & 25 (54.3) & 25 (54.3) & 27 (58.7) & 24 & 24 & 25 & 39--67 \\
cond-CDVAE & 46 & 17 (37.0) & 18 (39.1) & 19 (41.3) & 16 & 16 & 2 & 24--52 \\
TGDMat & 46 & 17 (37.0) & 17 (37.0) & 18 (39.1) & 15 & 15 & 13 & 24--52 \\
GemsDiff & 46 & 15 (32.6) & 15 (32.6) & 16 (34.8) & 17 & 15 & 14 & 20--46 \\
CDVAE & 46 & 12 (26.1) & 12 (26.1) & 12 (26.1) & 11 & 10 & 1 & 13--39 \\
CrystalGRW & 46 & 8 (17.4) & 8 (17.4) & 8 (17.4) & 6 & 6 & 4 & 6--28 \\
SymmCD & 46 & 7 (15.2) & 7 (15.2) & 8 (17.4) & 6 & 5 & 2 & 6--26 \\
\bottomrule
\end{tabular}
\end{table}

\begin{figure}[!htb]
\centering
\includegraphics[width=0.98\linewidth]{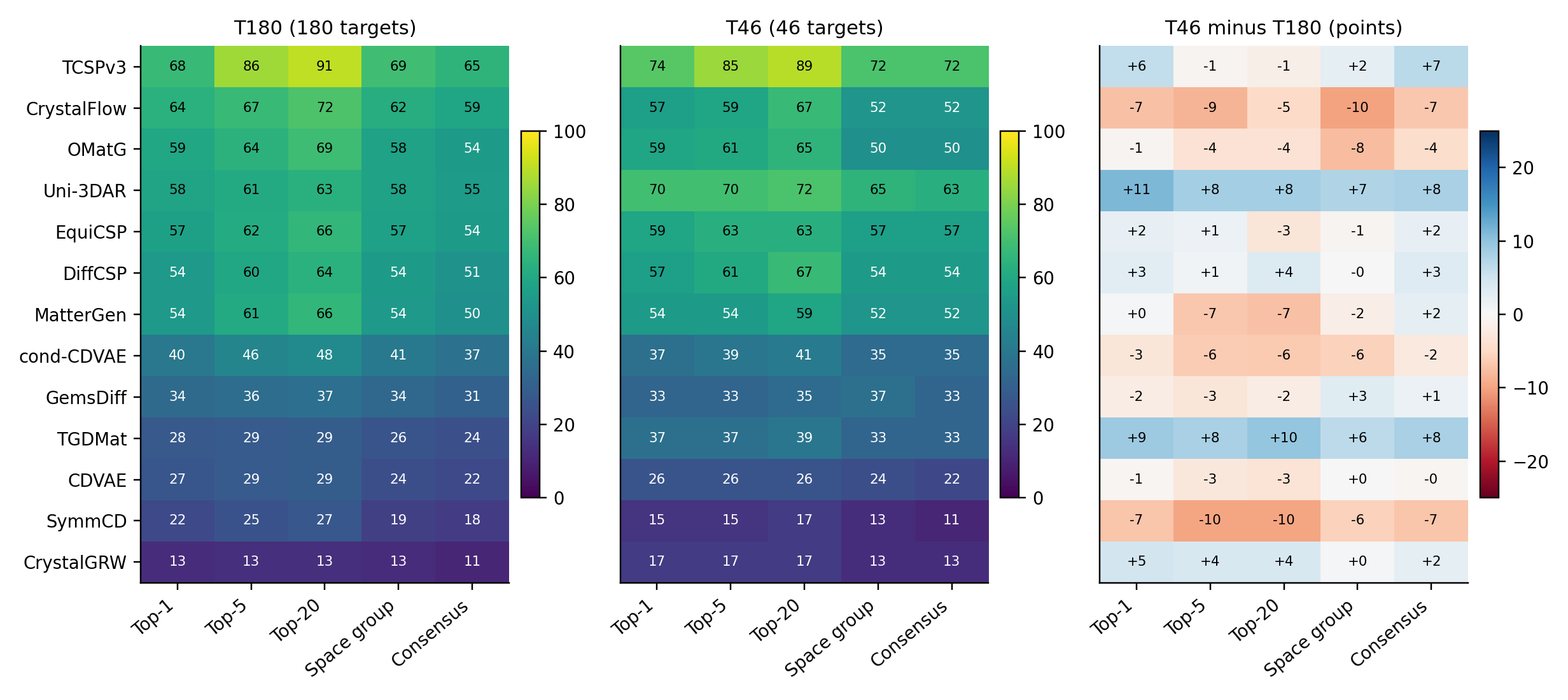}
\caption{Success rates (percent) of the 13 methods on T180 (left) and T46 (middle), and their
difference (right), in order of the T180 top-1 rate. Columns are the top-1, top-5 and top-20
StructureMatcher rates and the top-1 space-group and consensus rates. T46 shares 33 of its 46
targets with T180, so the two panels are not independent.}
\label{fig:heat-46}
\end{figure}

Figure~\ref{fig:heat-46} shows that the changes from T180 to T46 are mostly small and not
systematic: the median top-1 change is $+0$ points, the largest gains are
Uni-3DAR ($+11$) and TGDMat ($+9$), and the largest losses are CrystalFlow and SymmCD ($-7$).
Differences of this size correspond to one to five targets of 46 and are within the sampling noise
quoted above.

\subsection{Limited coverage beyond the template baseline}
\label{sec:complementary}

\begin{figure}[tbh!]
\centering
\includegraphics[width=0.8\linewidth]{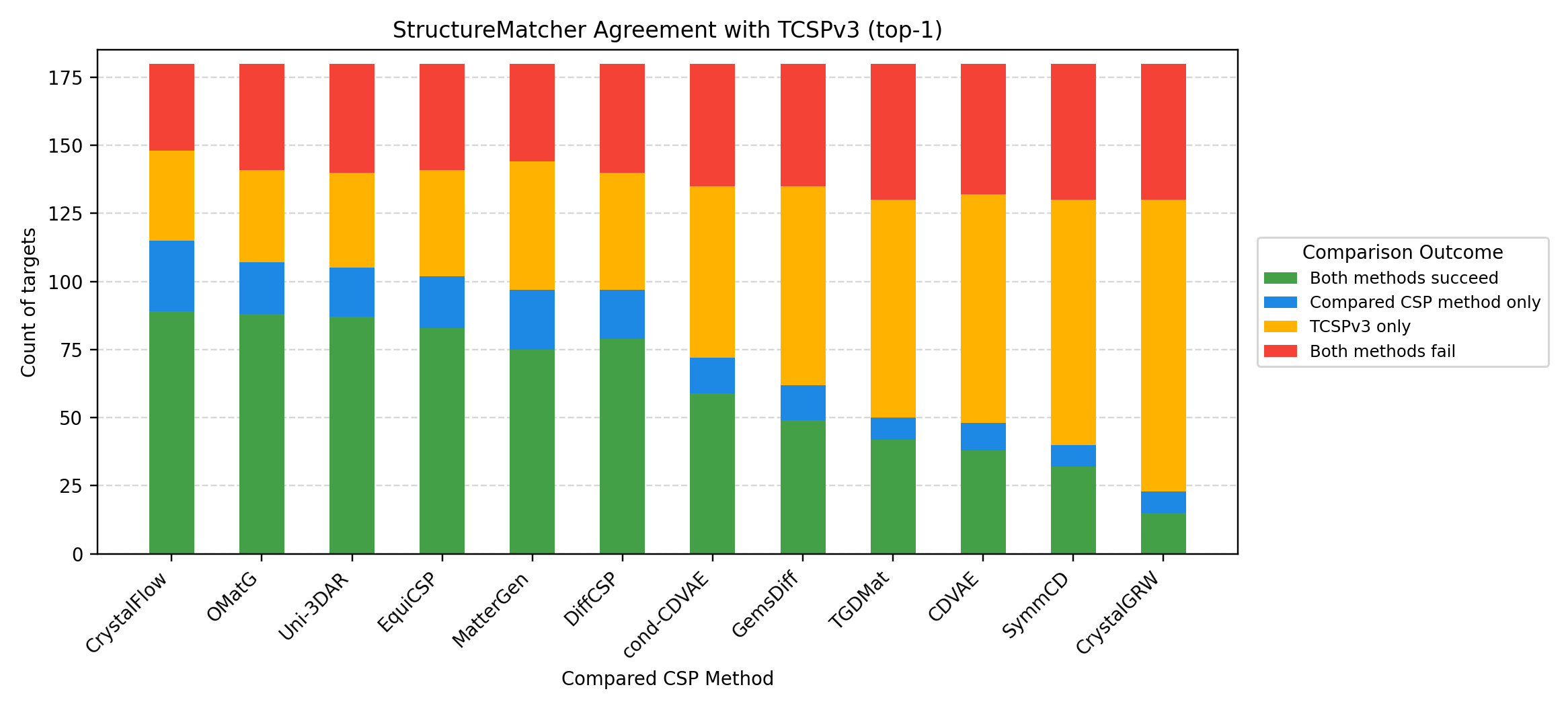}
\caption{StructureMatcher outcome decomposition (top-1) between TCSPv3 and each
generative model on the 180 targets of T180. Green: both succeed. Yellow:
TCSPv3 only. Blue: algorithm only. Red: neither. For every model, the green
segment exceeds the blue one, indicating that most of what generative models
predict correctly at top-1 is already recovered by template substitution; the
blue margin should not be read as template-independent prediction
(Section~\ref{sec:complementary}).}
\label{fig:tcsp-sm}
\end{figure}

\begin{figure}[tbh!]
\centering
\includegraphics[width=0.8\linewidth]{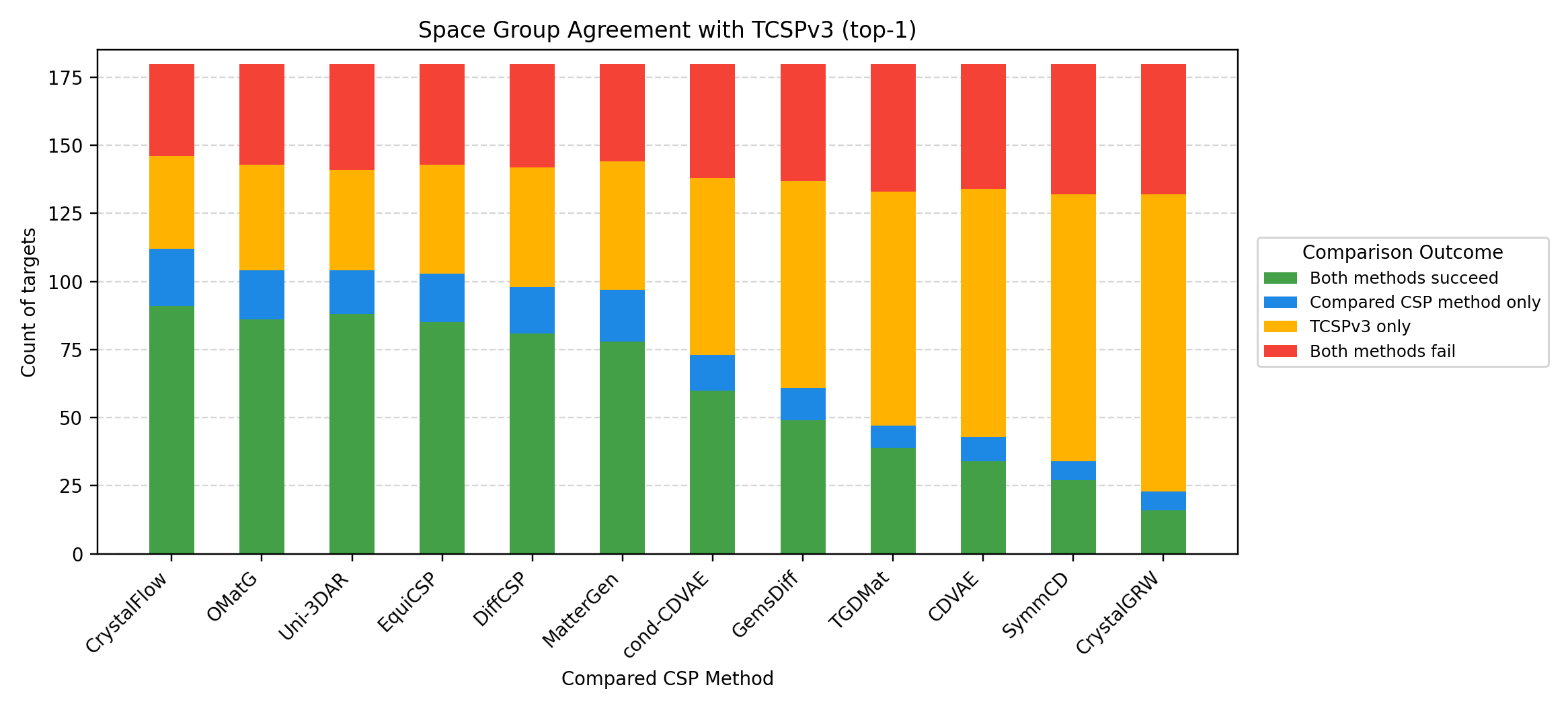}
\caption{Space-group match outcome decomposition (top-1) between TCSPv3 and each
generative model on T180, with segments as in Figure~\ref{fig:tcsp-sm}.}
\label{fig:tcsp-sg}
\end{figure}

\begin{figure}[tbh!]
\centering
\includegraphics[width=0.8\linewidth]{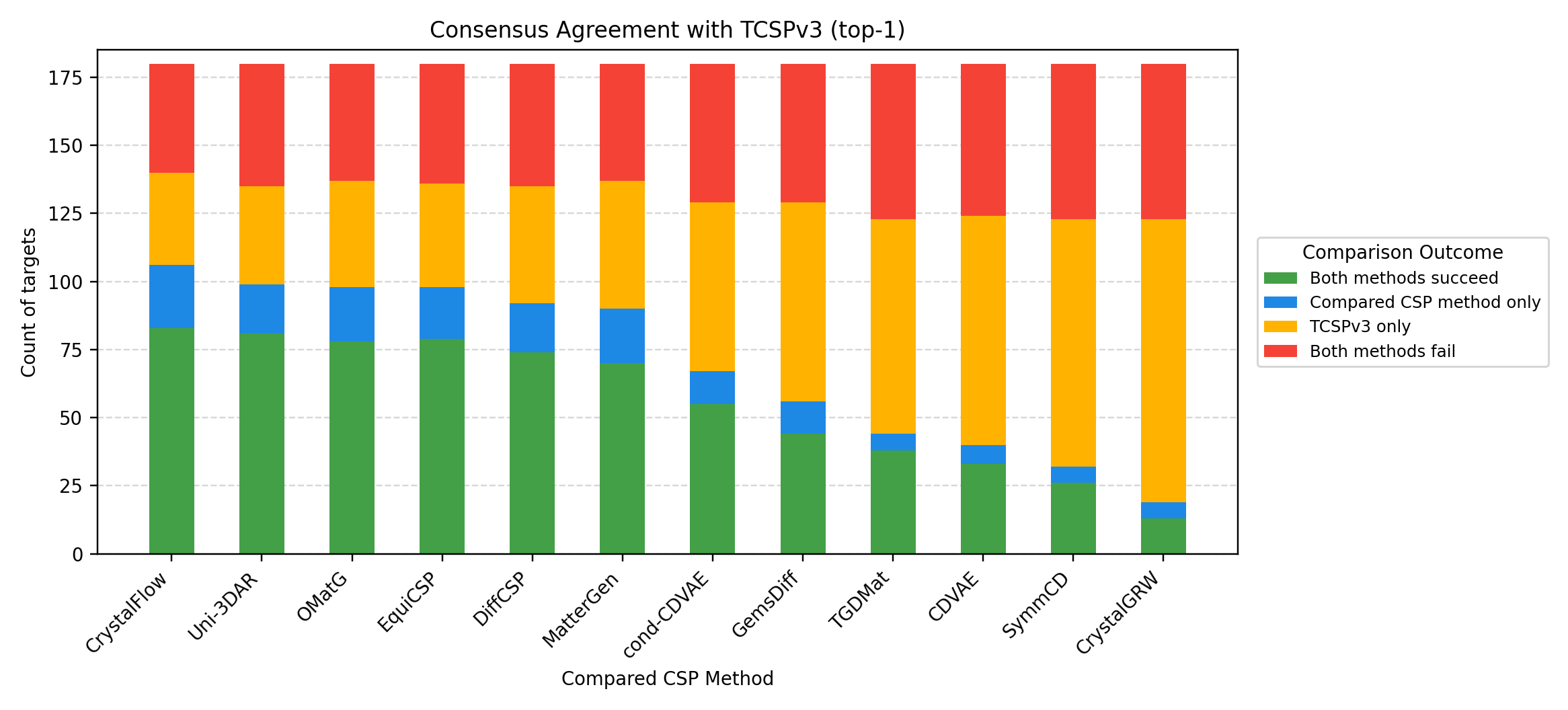}
\caption{Consensus outcome decomposition (top-1) between TCSPv3 and each
generative model on T180. Consensus requires simultaneous StructureMatcher and
space-group agreement for the top-1 candidate. TCSPv3 attains the highest
consensus rate (65.0\%).}
\label{fig:tcsp-consensus}
\end{figure}

Figures~\ref{fig:tcsp-sm}, \ref{fig:tcsp-sg} and~\ref{fig:tcsp-consensus} decompose each generative
model's top-1 performance against TCSPv3 into four categories per test
structure: both succeed, TCSPv3 only, algorithm only, and neither.

For every generative model in Figure~\ref{fig:tcsp-sm} the green segment is
larger than the blue one: the large majority of the targets a generative model
solves at top-1 are also solved by template substitution. CrystalFlow, for
example, succeeds on 115 targets, of which 89 are also solved by TCSPv3 and 26
are not. Across all twelve generative models, 33 of the 180 targets (18.3\%) are
solved at top-1 by at least one generative model while TCSPv3 fails; with
consensus as the criterion the number is 31 (17.2\%). The margin is therefore
not negligible at top-1, where TCSPv3's ranking of its own candidates is the
bottleneck, but it largely disappears when more candidates are allowed:
considering the top 20 candidates, only 8 targets (4.4\%) are solved by a
generative model and not by TCSPv3. At top-20, TCSPv3 alone solves 163 targets,
which is 6 more than the union of the twelve generative models (157).

This is a weakness for the \emph{de novo} framing, and we report it as one. The
motivating claim for generative CSP is that learning a distribution over
structures escapes the coverage limits of substitution on known prototypes. If
that were substantially true, we would expect a large algorithm-only segment that
persists as more candidates are allowed. We observe that the segment shrinks from
33 to 8 targets between top-1 and top-20, so most of it reflects a different
ordering of candidates, not structures that template substitution cannot reach.

The remaining margin cannot be taken at face value as evidence of
template-independence, for two reasons. First, TCSPv3 can fail for reasons
unrelated to prototype availability: the retrieval routes may not surface the
correct template among those carried forward, the element assignment may place
species on the wrong sites, or the radius-compatibility penalty may demote a
valid substitution. An algorithm-only success therefore demonstrates that the
two methods fail on different structures, not that no suitable prototype exists.

Second, we test this directly by partitioning every algorithm-only success at
top-1. For each such structure we determine whether a prototype sharing its
anonymized formula (sorted reduced stoichiometric coefficients, for example
AB$_2$) and its space group (spglib, \texttt{symprec}\,=\,0.01) exists elsewhere,
excluding the target itself. We check two corpora separately: the MP20 training
and validation data, which the generative models may have seen directly, and the
wider 731{,}293-structure library of TCSPv3 (with the MP20 test structures
removed), which the models have not seen but which represents what template
substitution can in principle draw on. A structure with a prototype in either corpus
is a failure of the template method, not a novelty; only a structure with no matching
prototype in either corpus supports a novelty claim.
Table~\ref{tab:prototype-availability} reports the result. Of 182
algorithm-only successes (target--model pairs, covering 33 distinct targets, listed in Supplementary Table~6), 157
(86.3\%) have a matching prototype in the MP20 training and validation data;
21 more have a match only in the wider library, and 4 (2.2\%) have no matching
prototype in either corpus. Those four instances are one target,
Ba$_6$Al$_{10}$ (space group 194), solved at top-1 by four models
(CrystalFlow, EquiCSP, OMatG and Uni-3DAR) and missed by TCSPv3 even at top-20.
Almost every algorithm-only success is therefore a TCSPv3 template-selection or ranking
failure rather than evidence of \emph{de novo} capability. The one
exception is the only algorithm-only success whose structure type is not
represented in either corpus under our definition, and it is the subject of one
of the case studies in Section~\ref{sec:case}. The definition is a
coarse proxy for a structural prototype, since it requires the same stoichiometry and space
group; the exception does have a parent with more species and the same framework in the MP20
data (Section~\ref{sec:case}).

\begin{table}[tb]
\centering
\caption{\textbf{Prototype availability among algorithm-only successes.}
For each generative method, we identify top-1 StructureMatcher successes on T180
that TCSPv3 does not achieve at top-1, and determine whether a matching
prototype is present in the MP20 training and validation data, only in the wider
731{,}293-structure TCSPv3 library, or in neither. A prototype match requires
the same sorted reduced stoichiometric coefficients and the same space group
(\texttt{symprec}\,=\,0.01); the target itself is excluded. Counts are
target--model pairs.}
\label{tab:prototype-availability}
\small
\begin{tabular}{lrrrr}
\toprule
& \textbf{Algorithm-only} & \multicolumn{3}{c}{\textbf{Prototype availability}} \\
\cmidrule(l){3-5}
\textbf{Method} & \textbf{successes} & \textbf{MP20 train/val} & \textbf{Wider library only} & \textbf{None} \\
\midrule
CrystalFlow  & 26 & 21 &  4 &  1 \\
MatterGen    & 22 & 19 &  3 &  0 \\
OMatG        & 19 & 15 &  3 &  1 \\
EquiCSP      & 19 & 16 &  2 &  1 \\
Uni-3DAR     & 18 & 15 &  2 &  1 \\
DiffCSP      & 18 & 15 &  3 &  0 \\
cond-CDVAE   & 13 & 12 &  1 &  0 \\
GemsDiff     & 13 & 13 &  0 &  0 \\
CDVAE        & 10 &  9 &  1 &  0 \\
TGDMat       &  8 &  7 &  1 &  0 \\
SymmCD       &  8 &  7 &  1 &  0 \\
CrystalGRW   &  8 &  8 &  0 &  0 \\
\midrule
\textbf{Total} & \textbf{182} & \textbf{157} & \textbf{21} & \textbf{4} \\
\bottomrule
\end{tabular}
\end{table}

\subsection{Prototype ablation: a causal test of template dependence}
\label{sec:ablation}

The preceding sections establish which models perform well and show that
outcome-level comparison against a template baseline cannot settle whether that
performance is prototype-dependent. We therefore intervene on the training
distribution directly.

\paragraph{Design.} We define a stoichiometric prototype family by reduced,
element-anonymous ratio, independent of element identity, ordering or space
group. For each of four families, AB$_2$ (1:2 binary), ABC$_2$ (1:1:2 ternary),
AB (1:1 binary) and ABC$_2$D$_6$ (1:1:2:6 quaternary), we remove every training
and validation structure of that family from the original MP-20 split (27{,}136
training and 9{,}047 validation structures), retrain from scratch on the
filtered data, and evaluate on the T180 targets of the removed family (23, 24,
19 and 11 targets; mp-3576, the HfGaCo$_2$ replacement, belongs to ABC$_2$).
Removal counts are given in Table~\ref{tab:ablation_removal}. The retained data
are not re-split, and each ablated model receives the same optimizer budget as its
baseline (318{,}000 steps for CrystalFlow, 106{,}000 for EquiCSP); the checkpoint
with the best validation loss is used for inference. Sampling, the 20 candidates
per target, and the ORB-v3 relaxation and energy ranking follow the benchmark
protocol of Section~\ref{sec:bench180}. The baseline is the corresponding model
trained on the full data, taken from the T180 campaign without re-running. A
target is \emph{degraded} if the baseline recovers it within the top 20 and the
ablated model does not.

We ablate CrystalFlow and EquiCSP, each on all four families. CrystalFlow is the strongest generative model
on T180 (Section~\ref{sec:bench180}), so its prototype dependence is a
conservative estimate for the paradigm. EquiCSP, which ranks fourth, is a second diffusion-based
model of lower rank, so the dependence can be compared across models of different strength.
Two further experiments were run for EquiCSP only. The first removes the AB$_3$ family (1:3 binary; 7 T180 targets). The second
tests the degeneracy explanation given below: it removes the ABC$_2$D$_6$ family and, in addition, every training and
validation structure whose framework (species ignored, supercells up to twice the cell allowed) matches that of any of the nine
perovskite-type ABC$_2$D$_6$ targets, and is evaluated on the 11 ABC$_2$D$_6$ targets. The test split is never filtered.

\begin{table}[!htb]
\centering
\caption{Training and validation structures removed in each ablation
experiment. Families are defined by reduced, element-anonymous stoichiometric
ratio. Ablated models were retrained from scratch on the filtered data; the
same removals apply to CrystalFlow and EquiCSP for the first four rows, whereas the AB$_3$ and
perovskite-framework experiments were run for EquiCSP only. The perovskite-framework removal comprises the
ABC$_2$D$_6$ family (1{,}551 training and 487 validation structures) and a further 1{,}268 and 410 structures
whose framework matches a perovskite-type target.}
\label{tab:ablation_removal}
\renewcommand{\arraystretch}{1.2}
\begin{tabular}{@{} l l r r r @{}}
\toprule
\textbf{Experiment} & \textbf{Prototype} &
\makecell{\textbf{Train}\\\textbf{removed}}  &
\makecell{\textbf{Validation}\\\textbf{removed}} &
\makecell{\textbf{Train}\\\textbf{remaining}} \\
\midrule
Remove AB$_2$        & All 1:2 binary        & 1{,}403 & 505   & 25{,}733 \\
Remove ABC$_2$       & All 1:1:2 ternary     & 4{,}026 & 1{,}351 & 23{,}110 \\
Remove AB            & All 1:1 binary        & 1{,}038 & 364   & 26{,}098 \\
Remove ABC$_2$D$_6$  & All 1:1:2:6 quaternary & 1{,}551 & 487   & 25{,}585 \\
Remove AB$_3$ & All 1:3 binary & 1{,}620 & 518 & 25{,}516 \\
Remove perovskite frameworks & \makecell[l]{ABC$_2$D$_6$ and\\perovskite frameworks} & 2{,}819 & 897 & 24{,}317 \\
\bottomrule
\end{tabular}
\end{table}

\paragraph{Removing a prototype family can cost a large share of targets, but
not uniformly.} Table~\ref{tab:ablation_results} gives the eight family arms and the two further EquiCSP experiments described below.
Removing the 1:2 binaries lowers CrystalFlow from 20 to 12 of 23 targets
recovered (8 degraded, 35\%) and EquiCSP from 17 to 14 (3 degraded, 13\%);
removing the 1:1:2 ternaries lowers CrystalFlow from 21 to 13 of 24 (9 degraded,
38\%) and EquiCSP from 21 to 13 (10 degraded, 42\%). In contrast, removing the 1:1 binaries barely changes either model: both
recover 15 of 19 targets against 16 at baseline, and only one target is lost. Removing the 1:1:2:6 quaternaries
does not lower either model: each recovers 10 of 11 targets against 8 at baseline, and neither loses a target on the structure match.
The consensus rate, which also requires the space group, follows the same
pattern (Table~\ref{tab:ablation_results}). Because the only variable that
changes is the presence of the family in training, and because the models are
retrained from scratch rather than fine-tuned, the losses are attributable to
prototype availability rather than to capacity, optimization or evaluation
differences. This is the form of evidence that observational novelty analyses
of generated outputs~\cite{negishi2026substitution} cannot supply. The
heterogeneity is equally informative: dependence is strong for the 1:2 binary
and 1:1:2 ternary families and weak for the 1:1 binaries. A plausible reason is a
degeneracy relation between prototypes: a structure with fewer distinct species can be obtained
from a retained structure with more species by merging species on equivalent sites, whereas the
reverse requires deciding which sites each species of the larger structure occupies, which a
model without a template cannot infer. A 1:1 binary query can therefore draw on retained
ternary and quaternary prototypes, but a ternary query cannot draw on binary ones. We checked
this on the retained MP-20 data. For each ablated target we counted the structures that remain
after removal of its family, have more distinct species than the target, and match its framework
(species ignored; supercells up to four times the target cell allowed). Eighteen of the 19 AB targets
have such a parent (median 77), against 12 of the 24 ABC$_2$ targets (median 0.5), which agrees
with the explanation: AB survives removal, and ABC$_2$ is the family that both models lose most of.
The ABC$_2$D$_6$ targets, which no model loses, also have parents (9 of 11; median 366), and these parents are the kind the explanation predicts: they are almost all five-species oxides of stoichiometry ABCDE$_6$ (for example CaLaMnMoO$_6$, SrCaNdSbO$_6$ and BaSrMgTeO$_6$), which reduce to the A$_2$BB$'$O$_6$ target when two species on equivalent sites are merged (Supplementary Table~8). Most of these
parents have a different space group from the target, which is expected, since merging species raises the symmetry
of a structure, so requiring the same space group and Wyckoff positions would undercount them (only 7 of 23 AB$_2$,
1 of 24 ABC$_2$ and none of the AB and ABC$_2$D$_6$ targets have such an exact parent). AB$_2$ does not follow the pattern for CrystalFlow:
all 23 targets have parents (median 38), yet CrystalFlow loses 8 of its 20 recovered targets. This does not invalidate the
explanation, because a generative model can fail on a target for many reasons besides the availability of a parent, and
parent availability is a necessary condition for recovery by this route, not a sufficient one. One further
possibility is chemical: a model conditioned on composition may not reach a parent whose elements differ strongly
from those of the query (among the AB$_2$ Laves phases, HfMn$_2$ and Be$_2$V are recovered and CaCd$_2$ and KAg$_2$ are lost,
although all four share ternary Laves parents such as Co$_3$SiMo$_2$). The parent count is a coarse proxy
for what a model can use, and with 11 to 24 targets per family we regard the degeneracy relation as a hypothesis
that is consistent with AB, ABC$_2$ and ABC$_2$D$_6$. We then tested the ABC$_2$D$_6$ case directly.

\paragraph{Removing the perovskite frameworks.} Removing the ABC$_2$D$_6$ family alone leaves EquiCSP's recovery of its 11 targets
unchanged, which the explanation attributes to the retained five-species perovskite parents. We therefore also removed every
training and validation structure with a perovskite framework (1{,}268 and 410 structures beyond the family) and retrained EquiCSP.
The structure match falls only from 8 to 7 of 11 targets (two lost, mp-19093 and mp-6389; mp-557387, missed at baseline, is now matched), but the consensus recovery falls from
8 to 3: five targets lose their consensus, the two with no matching candidate and three (mp-16307, mp-6603 and mp-12449)
whose recovered structure now has the wrong space group. Removing the parents therefore lowers the consensus recovery of these targets much more than the structure match, and in three of the five consensus
losses the model still produces a matching structure but with a different space group. This
rests on one retraining and 11 targets, so it is consistent with the explanation without establishing it. Removing the AB$_3$ family, a
family for which we have no parent analysis, costs EquiCSP one of the three AB$_3$ targets it recovers (of 7).

\begin{table}[!htb]
\centering
\caption{Ablation results on the T180 targets of each removed family.
\emph{Baseline fit} is the top-20 StructureMatcher recovery of the model trained
on the full data; \emph{ablated fit} and \emph{ablated consensus} are the
retrained model's top-20 recovery, and the same-candidate structure and
space-group agreement, on the same targets. \emph{Degraded} counts targets the
baseline recovers and the ablated model does not; percentages are relative to
all targets of the family. The last two EquiCSP rows are the additional experiments of
Table~\ref{tab:ablation_removal}; the perovskite-framework model is evaluated on the 11
ABC$_2$D$_6$ targets.}
\label{tab:ablation_results}
\renewcommand{\arraystretch}{1.2}
\small
\begin{tabular}{@{} l l r r r r r @{}}
\toprule
\textbf{Method} & \textbf{Experiment} &
\makecell{\textbf{Test}\\\textbf{targets}} &
\makecell{\textbf{Baseline}\\\textbf{fit}} &
\makecell{\textbf{Ablated}\\\textbf{fit}} &
\makecell{\textbf{Ablated}\\\textbf{consensus}} &
\makecell{\textbf{Degraded}} \\
\midrule
CrystalFlow & Remove AB$_2$        & 23 & 20 & 12 (52\%) & 12 (52\%) & 8 (35\%) \\
            & Remove ABC$_2$       & 24 & 21 & 13 (54\%) & 13 (54\%) & 9 (38\%) \\
            & Remove AB            & 19 & 16 & 15 (79\%) & 14 (74\%) & 1 (5\%) \\
            & Remove ABC$_2$D$_6$  & 11 & 8 & 10 (91\%) & 9 (82\%) & 0 (0\%) \\
\midrule
EquiCSP     & Remove AB$_2$        & 23 & 17 & 14 (61\%) & 14 (61\%) & 3 (13\%) \\
            & Remove ABC$_2$       & 24 & 21 & 13 (54\%) & 13 (54\%) & 10 (42\%) \\
            & Remove AB            & 19 & 16 & 15 (79\%) & 14 (74\%) & 1 (5\%) \\
            & Remove ABC$_2$D$_6$  & 11 & 8 & 10 (91\%) & 8 (73\%) & 0 (0\%) \\
            & Remove AB$_3$        & 7 & 3 & 2 (29\%) & 2 (29\%) & 1 (14\%) \\
            & Remove perovskite frameworks & 11 & 8 & 7 (64\%) & 3 (27\%) & 2 (18\%) \\
\bottomrule
\end{tabular}
\end{table}

\paragraph{Residual capacity for prototype-independent prediction.} A
substantial part of the family survives removal: even in the most damaging arms
(CrystalFlow AB$_2$ and ABC$_2$, EquiCSP ABC$_2$) about half of the targets are still
recovered, and the ablated models recover some targets that their baselines
missed (for example the ABC$_2$ target mp-1221342 under EquiCSP). A purely
template-based method would fail on every one of these targets by construction,
since with no prototype of the family available there is nothing to retrieve and
substitute. That the generators recover them indicates that they have acquired
structural regularities, plausibly local coordination preferences and element
compatibility, that transfer across prototype families. This is a positive,
quantified measure of generalization, measurable only under intervention. An
important qualification is that a family defined by stoichiometry is not the
same as a structure type: a ternary target may share its structure type with
retained quaternary or binary compounds, so the ablation removes the closest
prototypes but not every related one, and the residual is an upper bound on
prototype-free ability.

The two findings fit together rather than conflicting. Generative performance is
substantially prototype-dependent for the families that have been tested,
which is why Section~\ref{sec:complementary} finds so little coverage beyond the
template baseline: a model drawing largely on the same prototypes will largely
reach the same structures. The residual is what remains once that dependence is
subtracted, and it is a thin margin rather than the broad \emph{de novo}
capability that the literature claims.

\paragraph{Caveats.} The ablated targets are few (11 to 24 per family), so a
single target moves a rate by 4 to 9 percentage points and the differences between
CrystalFlow and EquiCSP on the same family are within that noise; we report counts
and do not rank the two models by the ablation. The EquiCSP baseline uses the
upstream final checkpoint, whereas the ablated EquiCSP models use the
validation-selected best checkpoint, so the baseline-to-ablated contrast mixes the effect of removal with a checkpoint-selection difference. Inference was run once per target.

\subsection{Case studies}
\label{sec:case}

\begin{figure}[!htb]
\centering
\includegraphics[width=0.95\linewidth]{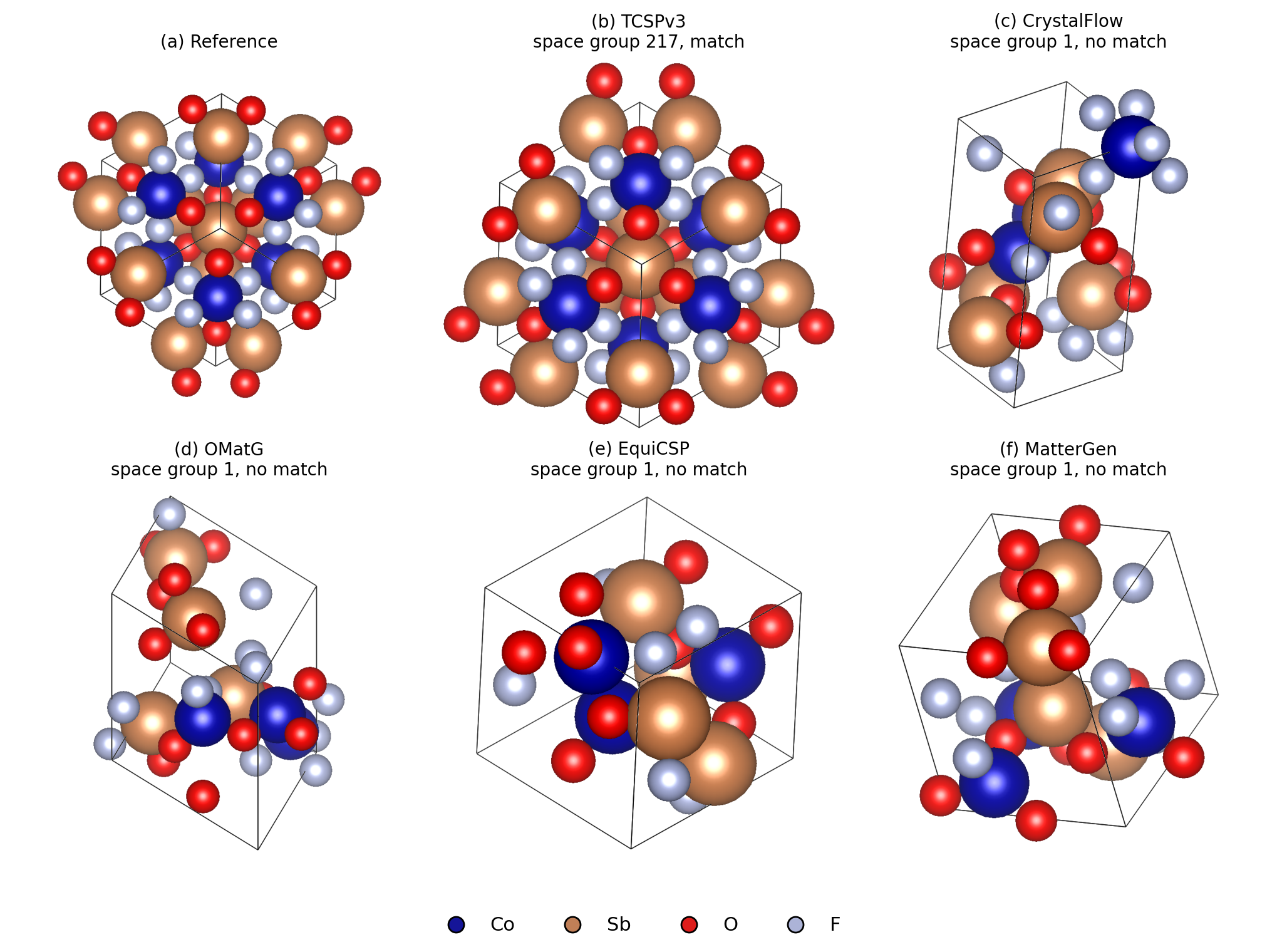}
\caption{Reference and top-1 predicted structures of Co$_3$Sb$_4$O$_6$F$_6$
(mp-1105290). (a) Reference. (b) TCSPv3. (c) CrystalFlow. (d) OMatG. (e) EquiCSP.
(f) MatterGen. Structures were drawn in VESTA with its default colours and radii, all as primitive cells (19 atoms), with
the reference reduced to its primitive cell, so that equivalent structures have the same cell and atom count. The space group
(\texttt{symprec}\,=\,0.1) and the StructureMatcher outcome of each candidate are
given in the panel titles.}
\label{fig:Co3Sb4O6F6}
\end{figure}

\begin{figure}[!htb]
\centering
\includegraphics[width=0.95\linewidth]{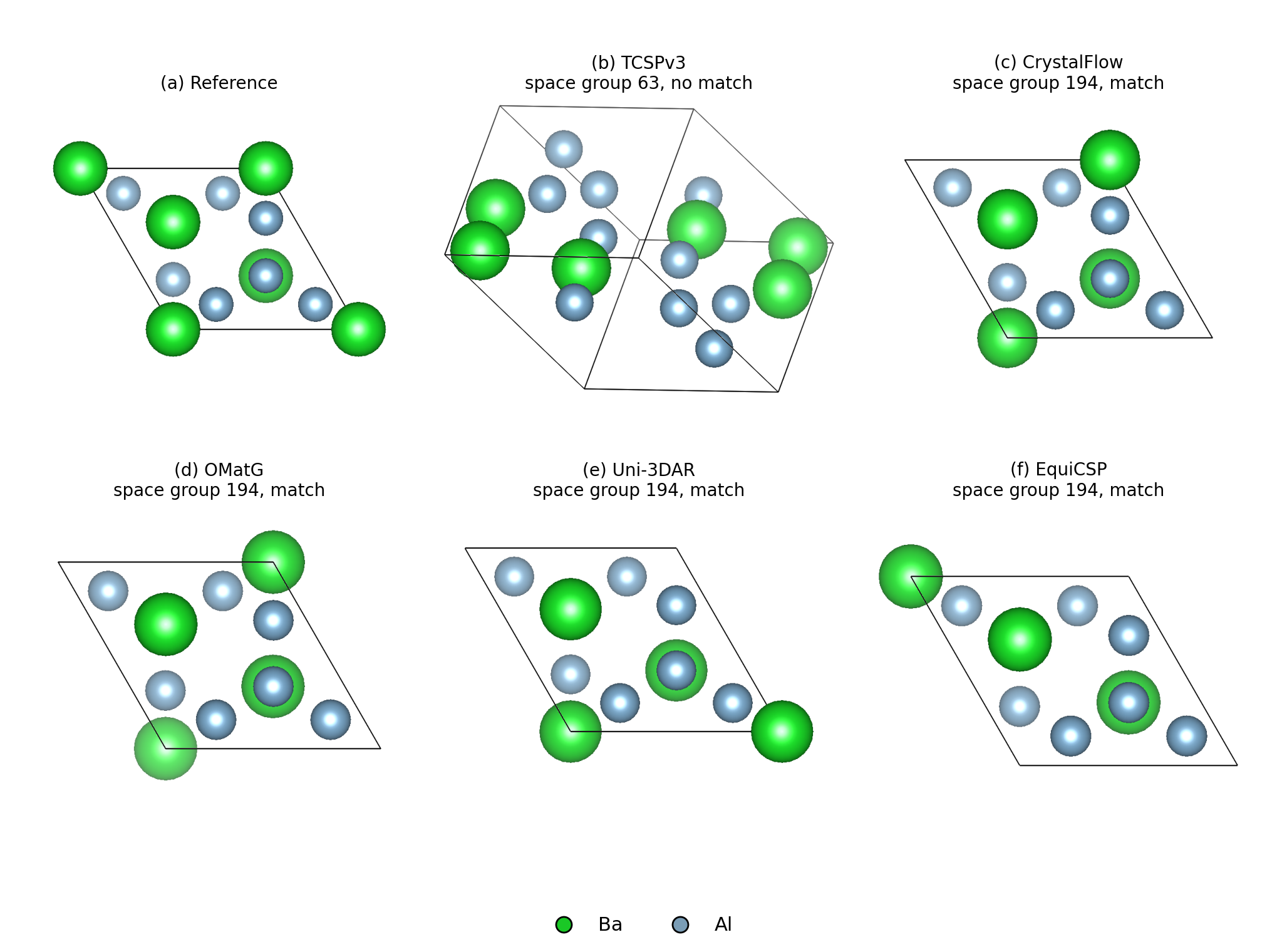}
\caption{Reference and top-1 predicted structures of Ba$_6$Al$_{10}$ (mp-261).
(a) Reference. (b) TCSPv3. (c) CrystalFlow. (d) OMatG. (e) Uni-3DAR.
(f) EquiCSP. Structures were drawn in VESTA with its default colours and radii, all as primitive cells (16 atoms). Panels (a) and (c)--(f) are viewed
along the hexagonal $c$ axis, and the orthorhombic TCSPv3 structure in (b) is shown in a different orientation; the panel titles give the space group (\texttt{symprec}\,=\,0.1) and the StructureMatcher
outcome.}
\label{fig:Ba6Al10}
\end{figure}

Two targets illustrate, from opposite directions, the margin described in
Section~\ref{sec:complementary}. Table~\ref{tab:case-study-distances} gives the
distance between the reference and the top-1 candidate of every method for both
targets, and Figures~\ref{fig:Co3Sb4O6F6} and~\ref{fig:Ba6Al10} show the structures.

\paragraph{Co$_3$Sb$_4$O$_6$F$_6$ (mp-1105290): template substitution succeeds
where generation fails.} This quaternary compound has a 19-atom primitive cell
(38 atoms in the conventional cell) and space group 217 ($I\bar{4}3m$). TCSPv3
matches it at top-1 with the correct space group and an energy distance of
0.000~eV/atom after relaxation with MatterSim, and it is closest to the
reference on the fingerprint and OFM distances (0.093 and 0.062). None of the
twelve generative models produces a matching structure among its 20 candidates,
and the top-1 candidate of every one of them has no symmetry (space group $P1$). Their top-1 candidates lie 0.037 to 0.099~eV/atom above the reference
energy, with fingerprint distances of 1.5 to 2.7 and OFM distances of 0.09 to
0.37. The library contains five prototypes with the same stoichiometry and
space group, and the MP20 data contain one, so the retrieval route has material
to work with. The pattern is consistent with the difficulty of assembling
multi-element structures without a prototype to anchor the arrangement.

\paragraph{Ba$_6$Al$_{10}$ (mp-261): generation succeeds where template substitution
fails.} This binary intermetallic has a 16-atom primitive cell and space group
194 ($P6_3/mmc$). CrystalFlow, EquiCSP, OMatG and Uni-3DAR each match it at
top-1 and recover the space group; their top-1 candidates are within
0.0001~eV/atom of the reference energy, with fingerprint distances of 0.12 to
0.14 and OFM distances of 0.010 to 0.011. TCSPv3 returns no matching structure among
its 11 candidates; its top-1 candidate, with space group 63, lies 0.025~eV/atom
above the reference, with a fingerprint distance of 1.54. It is the only
algorithm-only success for which neither the MP20 data nor the TCSPv3 library
contains a structure with the same stoichiometry and space group
(Table~\ref{tab:prototype-availability}), so a prototype-substitution account
cannot explain it directly. It is, however, reachable through the degeneracy relation of
Section~\ref{sec:ablation}: the MP20 training data contain one ternary structure,
Ba$_3$Al$_3$Ga$_2$ (mp-569218), with the same framework (space group $P6_3/mmc$, 16 atoms and the same
Wyckoff positions), and Ba$_3$Al$_5$ follows from it by merging Al and Ga on equivalent sites.
That four of the twelve models solve it is consistent with their exploiting this parent, but we
did not test it, which would require removing the parent and retraining; we also searched only the MP20 data
for such parents, not the wider library. We therefore do not read this structure as evidence
of prototype-independent prediction.

\begin{table}[tb]
\centering
\caption{\textbf{Distances from the reference structure for the two case-study
targets.} Each row compares the top-1 candidate of a method with the reference
primitive cell. Energy distance is the absolute difference in energy per atom
(eV/atom) between the MatterSim-relaxed candidate and the MatterSim-relaxed
reference. Chamfer distance and superposition RMSD (\AA{}) are computed on
Cartesian atom positions after the point-cloud registration of
CSPBenchMetrics, and fingerprint and OFM distances are Euclidean distances
between CrystalNN-fingerprint and orbital-field-matrix feature vectors. The
lowest value for each metric and composition is in bold. The Chamfer distance and
the registration RMSD depend on the cell origin and orientation of the
candidate, and should be read together with the fingerprint and OFM distances.}
\label{tab:case-study-distances}
\small
\setlength{\tabcolsep}{5pt}
\begin{tabular}{llrrrrr}
\toprule
\textbf{Composition} & \textbf{Method} & \textbf{Energy} & \textbf{Chamfer} & \textbf{Superposition} & \textbf{Fingerprint} & \textbf{OFM} \\
& & \textbf{distance} & \textbf{distance} & \textbf{RMSD} & \textbf{distance} & \textbf{distance} \\
\midrule
\multirow{13}{*}{Co$_3$Sb$_4$O$_6$F$_6$} & TCSPv3 & \textbf{0.000} & \textbf{4.044} & 1.485 & \textbf{0.093} & \textbf{0.062} \\
 & CrystalFlow & 0.070 & 6.700 & 1.763 & 2.431 & 0.267 \\
 & OMatG & 0.059 & 4.351 & 1.583 & 1.791 & 0.265 \\
 & Uni-3DAR & 0.088 & 7.599 & 1.681 & 2.711 & 0.368 \\
 & EquiCSP & 0.053 & 5.381 & 1.474 & 2.175 & 0.337 \\
 & MatterGen & 0.038 & 4.799 & 1.754 & 2.024 & 0.238 \\
 & DiffCSP & 0.067 & 4.594 & 1.869 & 1.826 & 0.138 \\
 & cond-CDVAE & 0.046 & 7.859 & 1.442 & 1.548 & 0.090 \\
 & GemsDiff & 0.037 & 5.626 & \textbf{1.412} & 2.112 & 0.220 \\
 & TGDMat & 0.065 & 4.271 & 1.480 & 2.598 & 0.244 \\
 & SymmCD & 0.067 & 7.378 & 1.578 & 1.972 & 0.271 \\
 & CrystalGRW & 0.072 & 10.636 & 1.828 & 2.210 & 0.260 \\
 & CDVAE & 0.099 & 6.244 & 1.992 & 2.169 & 0.318 \\
\addlinespace
\multirow{13}{*}{Ba$_6$Al$_{10}$} & TCSPv3 & 0.025 & 3.891 & 1.773 & 1.540 & 0.055 \\
 & CrystalFlow & \textbf{0.000} & \textbf{2.704} & 1.271 & 0.128 & \textbf{0.010} \\
 & OMatG & \textbf{0.000} & 3.293 & 1.623 & \textbf{0.124} & \textbf{0.010} \\
 & Uni-3DAR & \textbf{0.000} & 4.521 & \textbf{0.893} & 0.125 & \textbf{0.010} \\
 & EquiCSP & \textbf{0.000} & 3.384 & 1.635 & 0.141 & 0.011 \\
 & MatterGen & 0.043 & 5.347 & 1.951 & 1.936 & 0.154 \\
 & DiffCSP & 0.028 & 5.922 & 2.137 & 0.886 & 0.031 \\
 & cond-CDVAE & 0.044 & 5.750 & 1.839 & 1.691 & 0.044 \\
 & GemsDiff & 0.008 & 9.336 & 1.853 & 1.440 & 0.236 \\
 & TGDMat & 0.032 & 6.271 & 1.615 & 1.520 & 0.136 \\
 & SymmCD & 0.063 & 4.870 & 1.784 & 1.635 & 0.239 \\
 & CrystalGRW & 0.064 & 3.712 & 1.749 & 1.454 & 0.048 \\
 & CDVAE & 0.045 & 4.167 & 1.749 & 1.283 & 0.149 \\
\bottomrule
\end{tabular}
\end{table}

\FloatBarrier

\section{Discussion}
\label{sec:discussion}

\paragraph{What the benchmark shows.} On the 180 targets of T180, the template
baseline TCSPv3 solves 67.8\% (72.2\% with relaxing) of targets at top-1, 85.6\% (88.3\% with relaxing) at top-5 and
90.6\% at top-20, against 63.9\%, 67.2\% and 72.2\% (all CrystalFlow) for the best generative model
(Table~\ref{tab:main180}). It is therefore the strongest method at every candidate budget, with a lead of
7 targets at top-1 (15 with relaxing), 33 at top-5 (38 with relaxing) and 33 at top-20; ``with relaxing'' means that its candidates are
relaxed and ranked with ORB-v3, as those of the generative models are. The lead builds
with the budget because TCSPv3's candidate pool contains the correct template for many targets that it does not
rank first: 41 of its 58 top-1 failures are recovered within its top 20. Generative models, by contrast, gain
little from additional samples: CrystalFlow gains 15 targets between top-1 and
top-20, and Uni-3DAR 9. No generative model matches TCSPv3 at any budget, and TCSPv3 does so at a small fraction of their
computational cost. The advantage is therefore clearest when several candidates can be screened, and
it is already present when a single candidate is carried to DFT.

\paragraph{What the ablation does and does not establish.}
The intervention showed that removing a stoichiometric prototype family from
training caused CrystalFlow and EquiCSP to lose up to 48\% of the
targets they had previously recovered on that family, with the largest losses
for the 1:2 binary and 1:1:2 ternary families (Table~\ref{tab:ablation_results}). A further EquiCSP retraining without
the perovskite frameworks lowered its consensus recovery of the 1:1:2:6 targets from 8 to 3 of 11, which is
consistent with the parent-template reading. Read together with Section~\ref{sec:complementary}, the picture
is coherent: models that draw substantially on training prototypes will reach
substantially the structures those prototypes already reach, which is why the
coverage beyond the template baseline is thin once more than one candidate is
considered. It does not follow that generative models are equivalent to template
substitution. A template-based method has no residual at all once its templates are
removed, whereas the ablated generative CSP models still recovered 52--61\% of the targets in the most
damaging families, and that difference is real. Nor does the experiment show that the
remaining successes are unrelated to training data in any broader sense ---
structures from adjacent families may supply transferable motifs.
The ablation also points to a capability that generative models have and a substitution baseline lacks. Where retained structures with more species than the target exist (the 1:1 binaries, and the 1:1:2:6 quaternaries, which are reached from five-species ABCDE$_6$ oxides), removing the family cost the models few or no targets, whereas where such parents are scarce (the 1:1:2 ternaries) the losses were largest. This suggests that deep generative models can exploit templates of a different stoichiometry through the degeneracy relation, merging species on equivalent sites, without being told to; it is a property of learning a distribution over structures, not of the architecture. A template method has to be given this ability explicitly, for example by a cross-stoichiometry merge step, because plain substitution uses templates with the query's own number of species.

\paragraph{Relation to observational novelty analyses.} Our prototype partition
of the algorithm-only successes (Table~\ref{tab:prototype-availability}) is
convergent with the observational finding that 81--92\% of valid metastable
generated crystals are training duplicates or substitution-derived~\cite{negishi2026substitution}.
On T180, 157 of the 182 algorithm-only successes (target--model pairs; 86.3\%) have a prototype with
the same stoichiometry and space group in the MP20 training and validation data,
and 178 (97.8\%) have one in either that data or the TCSPv3 library. The two
analyses classify what models produce, and they agree that most successes that
look like prediction of a new structure have a close relative in known
structures. They differ in that ours restricts the question to targets that a
template baseline fails on, which is where a novelty claim would have to be
made. An ablation, which withholds inputs, answers the complementary question
of whether a model would still succeed had the substitution source been
withheld. The two approaches bound the question from opposite sides, and we
would encourage their joint use in future evaluations.

\paragraph{The residual.} Of the 182 algorithm-only successes, 4 have no
prototype of the same stoichiometry and space group in either corpus, and they are one target,
Ba$_6$Al$_{10}$, solved by four of the twelve generative models. It is the
only such case in T180, and the MP20 data contain a ternary structure with its framework,
Ba$_3$Al$_3$Ga$_2$, from which it follows by merging two species. It is therefore
consistent with the degeneracy relation that explains the ablation results, not evidence of
general capability. The definition of a prototype used for Table~\ref{tab:prototype-availability}
ignores Wyckoff positions and related stoichiometries, and whether the models would still
solve Ba$_6$Al$_{10}$ with this parent and its other structural neighbors removed is a question for a
further ablation.

\paragraph{Implications for benchmark design.} Aggregate match rate does not
measure novelty and should not be reported as if it did. A model can attain a
high match rate by covering the prototype distribution of the test set
thoroughly, which is useful but is precisely what template substitution already
does. Three practices follow from our results. First, every generative CSP paper
should report a template-substitution baseline on the same test set, and the
decomposition of Figure~\ref{fig:tcsp-sm} rather than aggregate rate alone. Most
current papers do neither, which is why a result as basic as the one in
Section~\ref{sec:complementary} --- that the overlap dominates the margin --- has
gone largely unremarked. Second, the number of candidates matters. The
algorithm-only margin is 33 targets at top-1 and 8 at top-20, so conclusions
about the added value of generative models depend on the budget at which they
are drawn. Third, prototype-controlled ablation or, where retraining is
infeasible, prototype-stratified reporting of match rates should become a
standard component of generative CSP evaluation. The cost is one retraining run
per prototype family, which is small relative to the cost of the models
themselves.

\paragraph{Interpretation of StructureMatcher results.} Structure-matching
tolerances vary widely across the literature, from \texttt{ltol}=0.3,
\texttt{stol}=0.5, \texttt{angle\_tol}=10 in the CDVAE
evaluation~\cite{xie2021crystal} to the stricter settings used
here~\cite{luo2024deep}. Rates are not comparable across settings, and
StructureMatcher agreement should be read jointly with space-group agreement,
whose value depends on the symmetry tolerance. On T180 the top-1 space-group rate
of CrystalFlow is 62.2\% at \texttt{symprec}\,=\,0.1 and 35.0\% at 0.01, and that
of TCSPv3 is 69.4\% and 67.8\%. A benchmark that reports a single space-group
number without its tolerance can therefore change the apparent standing of a generative model by
about 27 percentage points (CrystalFlow) against under 2 for a template method,
and structure matching alone can overstate fidelity to the symmetric structure. In both respects the
choice of metric and of tolerance changes the picture of the field.

\paragraph{Polymorphism and top-1 evaluation.} For compositions with multiple
known polymorphs we compare against each and retain the best match, which is
generous to all methods equally. Our primary metric is top-1, corresponding to
the practitioner carrying a single candidate to DFT; top-5 and top-20 results are
reported for comparability with the source literature and because, as shown
above, the ranking of methods is not the same at all budgets.

\paragraph{Limitations.} Our study has five limitations. First, the ablation
covers two models and four stoichiometric families (plus one further family and one framework removal for EquiCSP only), and each family contains only 7 to 24 targets, so the causal claim cannot be
generalized to the paradigm with confidence. Second, prototype families defined
by reduced stoichiometric ratio, or by stoichiometry and space group, are a coarse
proxy for structural prototype --- two structures with the same ratio may be
structurally unrelated, and two with different ratios may share a motif --- so the
measured dependence is an approximation whose direction of bias is not obvious.
Third, the screening of the test sets controls overlap with the MP20 training and
validation data only. The replacement targets are drawn from MPTS52, and the
CrystalGRW checkpoint was trained on a corpus we cannot
fully inspect, so overlap with that corpus cannot be excluded. Fourth, each method was
run once per target, so we report no sampling variance, and with 180 targets
one target equals 0.56 percentage points. Fifth, the methods are compared with their
own post-processing: generated structures are relaxed and ranked with ORB-v3,
whereas TCSPv3 candidates are ranked by their native chemical score, so the
top-1 comparison combines differences in generation and in ranking.

\paragraph{Energetic assessment.} The present evaluation is geometric,
comparing predicted and reference structures by matching and symmetry. It does
not assess whether predicted structures are thermodynamically competitive or
dynamically stable.

\section{Conclusion}

We investigate how much deep generative crystal structure prediction algorithms
add over template substitution based CSP algorithms, and answer it in two ways: by decomposing
predictions against a template baseline, and by intervening on the training
distribution directly.

The \emph{de novo} framing of current generative CSP algorithms is questionable. Our results
support a plainer reading: despite their sophistication, deep generative CSP models behave as
implicit template methods, and they perform below the explicit template method TCSPv3 at every
candidate budget, by 4 points at top-1 (8 points when TCSPv3 is relaxed as they are) and by 18 points at top-5 and top-20.
On the 180 targets of T180, the template baseline TCSPv3 solves 122 targets at
top-1 (67.8\%) and 163 at top-20 (90.6\%), against 115 (63.9\%) at top-1 and 130 (72.2\%) at top-20
for the best generative model. For every generative model evaluated, the large
majority of correct top-1 predictions are also recovered by TCSPv3. Thirty-three
targets (18.3\%) are solved at top-1 by some generative model where TCSPv3
fails, but only 8 (4.4\%) remain when 20 candidates are allowed, and TCSPv3
alone solves more targets at top-20 than the union of all twelve generative
models. Of the algorithm-only successes, 86.3\% have a prototype with the same
stoichiometry and space group in the MP20 training and validation data and 97.8\%
have one in either that data or the template library; the single exception, Ba$_6$Al$_{10}$,
has a ternary parent with the same framework in the MP20 data. A method that is orders of magnitude cheaper to run
recovers nearly the same set of structures. Whatever generative modeling
contributes here, it is not a substantial expansion of the reachable structure
space.

The ablation study explains the pattern. Removing entire stoichiometric prototype families from the
training data of a generative model and retraining from scratch caused the loss of
up to 48\% of the previously recovered targets of the removed family (CrystalFlow and
EquiCSP, T180), the loss being largest for the 1:2 binary and 1:1:2 ternary families and
negligible for the 1:1 binaries and absent for the 1:1:2:6 quaternaries, which establishes prototype dependence causally rather than by
association. Current generative CSP models are, to a first approximation,
implicit prototype libraries with softer edges than an explicit template library:
broader in what counts as a match, but drawing on the same underlying stock of
known structural motifs.

This is not to say that nothing remains. A small number of structures are still correctly
predicted after the removal of their prototype family, and one target, Ba$_6$Al$_{10}$, has no
same-stoichiometry prototype at all. Both are largely accounted for by the degeneracy relation: the
training data still contain structures with more species whose framework reduces to the target
when species are merged, and the models can use them, which plain substitution cannot do
without a designed merge step. This is a real property of generative models, but it is
the use of implicit templates, not prototype-independent prediction, and whether any residual
is independent of such parents is not established by our experiments.

We draw three recommendations. First, aggregate match rate should not be
reported as evidence of novelty; a model can score well by covering the test set's
prototype distribution thoroughly, which is precisely what substitution already
does. Prototype-controlled ablation, or prototype-stratified reporting where
retraining is infeasible, should become standard. Second, results should be
reported at several candidate budgets and with the symmetry tolerance stated,
since the comparison with a template baseline changes between top-1 and top-20
and between space-group tolerances. Third, architectural work should target the
residual directly (for example through prototype-aware augmentation) or through
training objectives enforcing invariance to elemental substitution across
prototype families, since enlarging the residual, rather than improving aggregate
match rate, is what would make generative CSP something substitution cannot
already do.

\section{Data and Code Availability}
The structures of the T180 and T46 test sets are drawn from the Materials Project
database~\cite{jain2013commentary}; their identifiers, the replacement
records of the MP20 audit, the top-20 predictions and per-target scores of all methods, the ablation split definitions (removed structure identifiers with file hashes), the candidates generated by every ablated model and the
analysis scripts are available at \url{https://github.com/usccolumbia/benchmark_deep_csp}, under the MIT license; the retrained ablation checkpoints are not released, and the arms can be
retrained from the split definitions. Performance metric code is
available at \url{https://github.com/usccolumbia/CSPBenchMetrics}.

\section{Contribution}
Conceptualization, J.H.; methodology, L.W., J.H., R.D., Y.F., M.M.; software, L.W.;
resources, J.H.; writing---original draft preparation, L.W., J.H., R.D., Y.F.;
writing---review and editing, J.H., M.M.; visualization, L.W.; supervision,
J.H.; funding acquisition, J.H.

\section*{Acknowledgement}
The research reported in this work was supported in part by the National Science
Foundation under grants 2110033, 2311202, and 2320292. The views, perspectives,
and content do not necessarily represent the official views of the NSF. The
authors gratefully acknowledge the computational resources provided by the Theia
high-performance computing cluster at the University of South Carolina, which is
supported by National Science Foundation Grant No. 2320292. We also acknowledge
the technical assistance and resources provided by Research Computing at the
University of South Carolina (RRID:SCR\_027488).

\bibliographystyle{unsrt}
\IfFileExists{references.bib}{\bibliography{references}}{}

\end{document}